\documentclass[pra,twocolumn,superscriptaddress,showpacs,10pt]{revtex4}
\usepackage{graphicx}
\usepackage{amssymb}

\begin{document}
\title{Geometric phase of an atom inside an adiabatic radio frequency potential}

\author{P. Zhang}
\affiliation{School of Physics, Georgia Institute of Technology,
Atlanta, Georgia 30332, USA}


\author{L. You}
\affiliation{School of Physics, Georgia Institute of Technology,
Atlanta, Georgia 30332, USA} \affiliation{Center for Advanced
Study, Tsinghua University, Beijing 100084, People's Republic of
China}

\date{\today}
\begin{abstract}
We investigate the geometric phase of an atom inside an adiabatic
radio frequency (rf) potential created from a static magnetic
field (B-field) and a time dependent rf field. The spatial
motion of the atomic center of mass is shown to give rise to a
geometric phase, or Berry's phase, to the adiabatically evolving
atomic hyperfine spin along the local B-field. This phase is found
to depend on both the static B-field along the semi-classical
trajectory of the atomic center of mass and an ``effective
magnetic field'' of the total B-field, including the oscillating rf
field. Specific calculations are provided for several recent atom
interferometry experiments and proposals utilizing adiabatic rf
potentials.
\end{abstract}
\pacs{03.65.Vf, 39.20.+q, 03.75.-b, 39.25.+k}
\maketitle

\section{Introduction}

Magnetic trapping is an important enabling technology for the active
research field of neutral atomic quantum gases. A variety of trap
potentials can be developed using magnetic (B-) fields with
different spatial distributions and time variations. For instance,
the widely used quadrupole trap and the Ioffe-Pritchard trap
\cite{IPT} are usually created with static B-fields, while the
time averaged orbiting potential (TOP) \cite{TOP} and time
orbiting ring trap (TORT) \cite{TORT,Kurn1} are created using
oscillating B-fields with frequencies larger than the effective
trap frequencies. Atom chips \cite{atomchip} have brought further
developments to magnetic trap technology, as they can provide
larger B-fields and gradients at reduced power-consumptions
or electric currents using micro-fabricated coils.
Today, magnetic trapping is a versatile tool used in many laboratories
around the world for controlling atomic spatial motion in regions
of different scales and geometric shapes, e.g., 3D or 2D traps,
double well traps, and storage ring traps
\cite{IPT,TOP,TORT,Kurn1,atomchip,Chapman,ring-2}.

Recently, magnetic traps based on adiabatic microwave
\cite{micro-theory,micro-experiment} and adiabatic radio frequency
(rf) potentials (ARFP)
\cite{Rf-collision,Rf-previous-theory,Rf-previous-hot,Rf-zhanghaichao,Rf-classical,Rf-quantum,Rf-experiment,
Rf-experiment2,Rf-experiment3,demarco,Rf-cooling,Rf-shanghai,Rf-ring,Rf-ring2,Rf-beyond,Rf-Zimmermann,Rf-combine}
have attracted considerable attention. An ARFP is typically
created with the combination of a static B-field and an rf field.
The idea for an ARFP has been around for some time
\cite{Rf-collision,Rf-zhanghaichao,Rf-previous-theory}, and
experimental demonstrations recently have been carried out for
confining both thermal \cite{Rf-previous-hot} and Bose condensed
atoms
 \cite{Rf-experiment,Rf-experiment2,Rf-experiment3,demarco}.
Further development of improved ARFP with atom chip technology
likely will assist in practical applications of atom
interferometry. For instance, a double well potential was
constructed recently using low order multi-poles capable of atomic
beam splitting while maintaining tight spatial confinement
\cite{Rf-classical}. Several interesting recent proposals outline
the construction of small storage rings with radii of the order
$1\mu$m
\cite{Rf-classical,Rf-quantum,Rf-combine,Rf-ring,Rf-ring2}, which
could become useful if implemented for atom Sagnac interferometry
\cite{Sagnac} setups on atom chips.

When a neutral atom is confined in a magnetic potential,
its hyperfine spin is assumed to follow adiabatically
the spatial variation of the B-field direction
during its spatial translational motion. As a result of this adiabatic
approximation, the center of mass motion for the atom
experiences an induced gauge field \cite{sun and other},
giving rise to a geometric phase (or Berry's phase)
to the atomic internal spin state \cite{berry,jorg}.
The effect of this geometric phase is widely known, and is first addressed
carefully in a meaningful way for atomic quantum gases
by an explicit calculation of the resulting geometric phase
in a static or a time averaged magnetic trap in Ref. \cite{Ho-Shenoy}.
Several important consequences are predicted to occur
for a magnetically trapped atomic condensate
in a quadrupole trap, a Ioffe-Pritchard
trap \cite{Ho-Shenoy,peng-prl}, or a TORT based storage ring
\cite{peng-pra}. To our knowledge, this geometric phase effect
has not been investigated in any detail for an atom inside an ARFP.

In a recent paper, we show that this geometric phase
causes an effective Aharonov-Bohm-type \cite{ABeffect} phase shift in
a magnetic storage ring based atom interferometer
\cite{peng-pra}. In addition, our studies
imply that the spatial fluctuation of the geometric
phase can lead to a reduction of the visibility of the interference
contrast. In view of this, we decided to carry out this
study as reported here for the atomic geometric phase in an ARFP
in order to shed light on the proposed high precision
atom Sagnac interference experiment \cite{Sagnac}.

Analytical derivations for this study at some places
become rather tedious and complicated.
We therefore first will summarize
our major results here for readers who may not be
interested in the intricate details.
We find that the geometric phase in an ARFP generally takes a more
complicated form in comparison to the case of a static trap or a
time averaged trap. In an ARFP, this phase factor is
found to be determined by the trajectory of
the time independent component of the trap field as well as an
``effective B-field" that depends on the total
B-field. In contrast to the earlier result found for
a static trap or a time averaged trap \cite{Ho-Shenoy,peng-pra},
the final result turns out to be not expressible
as a functional of the trajectory for the direction of the total
B-field in the parameter space.

This paper is organized as follows. In sec. II, we generalize the
semi-classical approach as outlined in Ref. \cite{Rf-classical}
for the operating principle of an ARFP to a form more convenient
for discussing the geometric phase. Section III parallels that of
sec. II by reformulating a full quantum theory for discussing the
geometric phase inside an ARFP \cite{Rf-quantum}. The explicit
expression for the geometric phase inside an ARFP is given in a
readily adaptable form for specifical calculations. In sec. IV, we
discuss the effect of the geometric phase in several types of ARFP
recently proposed for atomic splitters and storage rings
\cite{Rf-classical,Rf-quantum,Rf-combine,Rf-ring,Rf-ring2} and
beam splitters \cite{Rf-experiment,Rf-classical}. Finally,
concluding remarks are given in sec. V.

\section{A semi-classical approach}

In this section, we provide a semi-classical formulation
for calculating the atomic geometric phase inside an ARFP.
The semi-classical working principle for an ARFP is
described in Ref. \cite{Rf-classical},
although only for the special case when the atomic center
of mass is assumed at a fixed location.
In order to calculate the geometric phase, our formulation
allows for the explicit
consideration of atomic center of mass motion classically.
In our approach, the geometric phase is obtained naturally,
and the validity conditions for both the adiabatic
and the rotating wave approximations are clearly shown
for an ARFP.

Inside an ARFP \cite{Rf-classical}, the total
B-field ${\vec B}(\vec{r},t)$ is the sum of
a static field component ${\vec B}_{s}(\vec{r})$
and an oscillatory rf field ${\vec B}_{o}(\vec{r},t)$,
which conveniently is expressed as
\begin{eqnarray}
{\vec B}_{o}(\vec{r},t)={\vec
B}^{(a)}_{\rm{rf}}(\vec{r},t)\cos(\omega t)+{\vec
B}^{(b)}_{\rm{rf}}(\vec{r},t)\cos(\omega t+\eta).
\label{BO}
\end{eqnarray}
where $\vec{r}$ is the spatial position vector of the atom,
$\omega$ is frequency of the rf field, and $\eta$ is
a relative phase factor.

In this section, we will assume that the atomic
spatial motion is pre-determined, i.e., $\vec{r}(t)$ is
given (as a slowly varying function of time $t$).
For weak B-fields, the system Hamiltonian is simply the linear Zemman interaction
\begin{eqnarray}
H(t)=g_{F}\mu_{B}{\vec F}\cdot{\vec B}[\vec{r}(t),t],
\label{H}
\end{eqnarray}
where $g_{F}$ is the corresponding \textit{Lande} g-factor and
$\mu_{B}$ denotes the Bohr magneton.
$\hbar=1$ is assumed.

For a static or a time averaged magnetic trap,
the Hamiltonian (\ref{H})
varies slowly over time scales of the Larmor precession
of the atomic spin in the total B-field.
During the effectively slow trapped motion, the atomic hyperfine spin
is assumed to be fixed at the instantaneous eigenstate
of the Hamiltonian (\ref{H}).
The geometric phase then can be calculated straightforwardly
from the variation of the B-field direction
in the parameter space \cite{Ho-Shenoy,peng-pra}.

In an ARFP, the situation is more complicated.
Although the variation of
$\vec{B}_{s}[\vec{r}(t)]$ remains much slower than
the Larmor precession, the rf frequency $\omega$
usually is assumed to be nearly resonant with the precession frequency.
Thus, the Hamiltonian (\ref{H}) contains both fast and
slow time varying components, making the direct calculation
of the geometric phase a more involved task.
In the following, we will proceed step by step,
clarifying the various approximations adopted along the way.

To understand the working principle for an ARFP,
we first decompose the Hamiltonian $H(t)$ (\ref{H})
into the following form
\begin{eqnarray}
H(t)=H_{s}[\vec{r}(t)]+H_{+}[\vec{r}(t)]e^{-i\omega
t}+H_{-}[\vec{r}(t)]e^{i\omega t},
\end{eqnarray}
where $H_{s}$ and $H_{\pm}$ are all slow varying functions
of time and are given by
\begin{eqnarray}
H_{s}[\vec{r}(t)]&=&g_{F}\mu_{B}{\vec F}\cdot{\vec
B}_{s}[\vec{r}(t)],\nonumber\\
H_{+}[\vec{r}(t)]&=&\frac{1}{2}g_{F}\mu_{B}{\vec F}\cdot\left({\vec
B}^{(a)}_{\rm{rf}}[\vec{r}(t)]+e^{-i\eta}{\vec B}^{(b)}_{\rm{rf}}[\vec{r}(t)]\right), \nonumber\\
H_{-}[\vec{r}(t)]&=&H_{+}^{\dagger}[\vec{r}(t)].
\end{eqnarray}

$H_{s}$ is diagonal in the spin angular
momentum basis defined along the local direction
of the static B-field $\vec B_{s}[\vec{r}(t)]$.
The eigenstate takes the familiar form
$|m_{F}[\vec{r}(t)]\rangle_s$, quantized along the direction
of $\vec B_{s}[\vec{r}(t)]$,
with the eigenvalue $m_F\left|B_{s}[\vec{r}(t)]\right|$
for $\vec{B}_{s}[\vec{r}(t)]\cdot\vec{F}$ and
$m_F\in [-F,F]$, in analogy with the
usual case of the z-quantized representation result of
$F_{z}|m_F\rangle_z=m_F|m_F\rangle_z$.

Next we introduce a unitary transformation
\begin{eqnarray}
U(t)=\sum_{m_F=-F}^{F}|m_F\rangle_z\,_s\langle
m_{F}[\vec{r}(t)]|e^{im_F\kappa\omega t},
\label{U}
\end{eqnarray}
with $\kappa={\rm sign}(g_{F})$ for the rotating wave approximation.
The quantum state
in the interaction picture $|\Psi(t)\rangle_{I}=U(t)|\Psi(t)\rangle$ defined by $U(t)$ is
governed by the Schroedinger equation
$i\partial_{t}|\Psi(t)\rangle_{I}=H_{I}(t)|\Psi(t)\rangle_{I}$,
with the Hamiltonian in the interaction picture given by
\begin{widetext}
\begin{eqnarray}
H_{I}(t)&=&UHU^{\dagger}+i(\partial_{t}U)U^{\dagger}\nonumber\\
&=&\sum_{m=-F}^{F}m\kappa\Delta[\vec{r}(t)]|m\rangle_z\,_z\langle
m|-i\sum_{m,n=-F}^{F}|m\rangle_z\,_s\langle
m[\vec{r}(t)]|\frac{d}{dt}|n[\vec{r}(t)]\rangle_s\,_z\langle
n|e^{i(m-n)\kappa\omega t}\nonumber\\
&&+\sum_{m=-F+1}^{F}\left(h^{(+)}_{m}[\vec{r}(t)]|m\rangle_z\,_z\langle
m-1|+h^{(-)}_{m}[\vec{r}(t)]|m\rangle_z\,_z\langle m-1|e^{2i\kappa\omega
t}+h.c.\right)\nonumber\\
&&+\sum_{m=-F}^{F}\left(h_{m}[\vec{r}(t)]|m\rangle_z\,_z\langle m|e^{i\kappa\omega t}
+h.c.\right),\ \ \ \
\label{HIprecise}
\end{eqnarray}
\end{widetext}
where the time dependent parameters are defined as
\begin{eqnarray}
\Delta[\vec{r}(t)]&=&\mu_{B}|g_{F}\vec{B}[\vec{r}(t)]|-\omega,\nonumber\\
 h^{(\pm)}_{m}[\vec{r}(t)]&=&\,_s\langle
m[\vec{r}(t)]|H_{\pm}[\vec{r}(t)]|(m-1)[\vec{r}(t)]\rangle_s,\ \ {\rm and} \label{HI}\\
h_{m}[\vec{r}(t)]&=&\,_s\langle
m[\vec{r}(t)]|H_{\pm}[\vec{r}(t)]|m[\vec{r}(t)]\rangle_s.\nonumber
\end{eqnarray}

The above result is obtained easily if we note that
the matrix element $\,_s\langle m[\vec{r}(t)|H_{\pm}(t)|m'[\vec{r}(t)\rangle_s$
is non-zero only when $m-m'=0,\pm 1$. So far, we have always assumed that
$|m[\vec{r}(t)]\rangle_s$ is a single valued function of the
atomic position $\vec{r}$. A careful examination shows that
the eigenstate $|m[\vec{r}(t)]\rangle_s$ cannot be determined
uniquely because of the presence of the $U(1)$ gauge freedom for selecting
a local phase factor $\exp\{i\phi[\vec{r}(t)]\}$,
which consequently affects the resulting expressions for
$h_{m}^{(\pm)}(t)$ and $\,_s\langle m[\vec{r}(t)]|d/dt|m'[\vec{r}(t)]\rangle_s$.

The rotating wave approximation neglects of the
oscillating terms proportional to $e^{im\omega t}$
($m\neq 0$) in the Hamiltonian $H_I$ (\ref{HIprecise}).
The error for this approximation is
estimated easily from a time dependent perturbation calculation.
The sufficient condition for its validity requires that all factors such as
$\int_{0}^{t}dt'h_{m}(t')\exp[i\kappa\omega t']$,
$\int_{0}^{t}dt'h_{m}^{(-)}(t')\xi_{m,m-1}(t')\exp[i\kappa(2\omega+\Delta)
t']$, and $\int_{0}^{t}dt'\langle
m[\vec{r}(t')]|d/dt'|n[\vec{r}(t')]\rangle\xi_{mn}(t')\exp[i(m-n)\kappa(\omega+\Delta)
t']$ are negligible, where
\begin{eqnarray}
\xi_{mn}(t)&=&\exp\left[\int_{0}^{t}dt'\,_s\langle
m[\vec{r}(t')]|\frac{d}{dt'}|m[\vec{r}(t')]\rangle_s\right]\times\nonumber\\
&& \exp\left[-\int_{0}^{t}dt'\,_s\langle
n[\vec{r}(t')]|\frac{d}{dt'}|n[\vec{r}(t')]\rangle_s\right].
\end{eqnarray}
Thus, the gauge independent factors
$h_{m}^{(-)}\xi_{m,m-1}$, $\langle
m_{s}|d/dt|n_{s}\rangle\xi_{mn}$ , $h_{m}$, and $\Delta$
should all vary slowly with time and
with the modulus of their amplitudes much less than $\omega$.

The effective Hamiltonian in the
interaction picture under the rotating wave approximation then becomes
\begin{eqnarray}
H^{(I)}_{\rm eff}(t)&=&\mu_{B}g_{F}\vec{F}\cdot\vec{B}^{\rm eff}[\vec{r}(t)]\nonumber\\
&&-i\sum_{m=-F}^{F}|m\rangle_z\,_s\langle
m[\vec{r}(t)]|\frac{d}{dt}|m[\vec{r}(t)]\rangle_s\,_z\langle m|,\
\ \ \ \ \label{HIeff}
\end{eqnarray}
where the first term resembles a coupling between the
atomic spin and an ``effective B-field" $\vec{B}^{\rm
eff}(\vec{r})$, whose components in real space are given by
\begin{eqnarray}
B_{x}^{\rm eff}(\vec{r})&=&{\rm Re} \left[\frac{2\,_s\langle
m(\vec{r})|H_{\pm}(\vec{r})|(m-1)(\vec{r})\rangle_s}
{\mu_{B}g_{F}\sqrt{(F+m)(F-m-1)}}\right],\nonumber\\
B_{y}^{\rm eff}(\vec{r})&=&-{\rm Im}\left[\frac{2\,_s\langle
m(\vec{r})|H_{\pm}(\vec{r})|(m-1)(\vec{r})\rangle_s}
{\mu_{B}g_{F}\sqrt{(F+m)(F-m-1)}}\right], \ \ {\rm and} \nonumber \\
B_{z}^{\rm eff}(\vec{r})&=&\left|\vec{B}_{s}(\vec{r})\right|
-\frac{\omega}{\mu_{B}|g_{F}|}\ .\ \ \ \label{BI}
\end{eqnarray}
Clearly, the $x$- and $y$-components of the effective field
$\vec{B}^{\rm eff}(\vec{r})$ depend on the explicit form of the
eigenstate $|m(\vec{r})\rangle_s$. In fact, it easily can be seen
that different choices of the local phase factor for the
$|m(\vec{r})\rangle_{s}$ actually lead to different values of
$\vec{B}^{\rm eff}(\vec{r})$ related to each other through
$\vec{r}$-dependent rotations in the $x$-$y$ plane.

In practice, the eigenstate $|n(\vec{r})\rangle_s$ and the
effective field $\vec{B}^{\,\rm eff}$ can sometimes be constructed
more simply, as in Ref. \cite{Rf-classical}. For any spatial
position $\vec{r}$, we first choose a rotation
$R[\hat{m}(\vec{r}),\chi(\vec{r})]$ along the axis
$\hat{m}(\vec{r})$ with an angle $\chi(\vec{r})$ that satisfies
$R[\vec{n}(\vec{r}),\chi(\vec{r})]\vec{B}_{s}(\vec{r})=|\vec{B}_{s}(\vec{r})|\hat{e}_{z}$.
It is then easy to show that the eigenstate
$|n[\vec{r}(t)]\rangle_s$ can be chosen as
\begin{eqnarray}
|n(\vec{r})\rangle_s=\exp\left[i\vec{F}\cdot\hat{m}(\vec{r})\chi(\vec{r})\right]|n\rangle_z.
\label{nss}
\end{eqnarray}
Unfortunately, the choice for $R$ is not unique in a given static
field $\vec{B}_{s}(\vec{r})$, an analogous result to the $U(1)$
gauge freedom for the the egienstate $|n(\vec{r})\rangle_{s}$.
Corresponding to the choice (\ref{nss}) given above for
$|n(\vec{r})\rangle_s$, the unitary transformation $U$ defined in
(\ref{U}) would become
\begin{eqnarray}
U(t)=\exp(-iF_{z}\omega t)\cdot\exp\left[-i\vec{F}\cdot\hat{m}(\vec{r})\chi(\vec{r})\right],
\end{eqnarray}
and the transverse components of the ``effective B-field"
given by $B^{\,\rm
eff}_{x,y}(\vec{r})=\overline{B}_{x,y}(\vec{r})/2$ \cite{Rf-classical}
with
\begin{eqnarray}
\overline{\vec
B}(\vec{r})&=&R[\hat{m}(\vec{r}),\chi(\vec{r})]\vec B_{\rm
rf}^{(a)}(\vec{r})\nonumber\\
&&+R[\hat{e}_{z},-\kappa\eta]R[\hat{m}(\vec{r}),\chi(\vec{r})]\vec B_{\rm
rf}^{(b)}(\vec{r}).
\end{eqnarray}

In earlier discussions of an ARFP \cite{Rf-classical,Rf-quantum},
the atomic internal state is assumed uniformly to
remain adiabatically in a certain eigenstate of the first term of $H^{(I)}_{\rm eff}(t)$.
To fully appreciate this adiabatic approximation and to calculate
the geometric phase, we expand $|\Psi(t)\rangle_{I}$ into the
instantaneous eigenstate basis $|n[\vec{r}(t)]\rangle_{\rm eff}$
quantized along the direction of the effective B-field $\vec{B}^{\rm eff}$
according to $|\Psi(t)\rangle_{I}=\sum_{n}C_{n}(t)|n[\vec{r}(t)]\rangle_{\rm eff}$.
The first term of $H^{(I)}_{\rm eff}(t)$ is simply the
effective Zemman interaction between the atomic hyperfine spin and
the effective B-field.
The corresponding Schroedinger equation for
the Hamiltonian $H^{(I)}_{\rm eff}(t)$ of (\ref{HIeff}) then becomes
\begin{eqnarray}
i\frac{d}{dt}C_{n}(t)=[\epsilon_{I}^{(n)}(t)+\nu_{nn}(t)]C_{n}(t)+\sum_{m\neq
n}\nu_{nm}(t)C_{m}(t), \nonumber
\end{eqnarray}
with
\begin{eqnarray}
\epsilon_{I}^{(n)}(t)&=&n\mu_{B}g_{F}|\vec{B}^{\rm eff}(t)|,\nonumber\\
\nu_{pq}(t)&=&-i\sum_{l}\,_{\rm eff}\langle p[\vec{r}(t)]|l\rangle_z\nonumber\\
&&\,_s\langle
l[\vec{r}(t)]|\frac{d}{dt}|l[\vec{r}(t)]\rangle_s\,_z\langle
l|q[\vec{r}(t)]\rangle_{\rm eff} \nonumber\\
&&-i\,_{\rm eff}\langle
p[\vec{r}(t)]|\frac{d}{dt}|q[\vec{r}(t)]\rangle_{\rm
eff}.
\end{eqnarray}

Under the adiabatic approximation,
the atomic internal state remains
in a given eigenstate $|n[\vec{r}(t)]\rangle_{\rm eff}$ with
transitions to states $|m[\vec{r}(t)]\rangle_{\rm eff}$
($m \neq n$) being negligibly small.
Thus, the transition probability, as estimated from
the first order perturbation theory,
\begin{eqnarray}
\int_{0}^{t}dt'\nu_{nm}(t')e^{i\int_{0}^{t'}dt''[\epsilon_{I}^{(n)}(t'')+\nu_{nn}(t'')
-\epsilon_{I}^{m}(t'')-\nu_{mm}(t'')]} \nonumber
\end{eqnarray}
should be much less than one.
As before, we find the sufficient condition for the validity of
the adiabatic approximation is given by
\begin{eqnarray}
\frac{|\nu_{mn}(t)|}{|\epsilon_{I}^{m}(t')-\epsilon_{I}^{(n)}(t')|}\ll 1,
\end{eqnarray}
provided that
$\nu_{mn}(t)\exp[i\int_{0}^{t'}[\nu_{nn}(t'')-\nu_{mm}(t'')]dt']$,
which is independent of the local phase factor for
$|n[\vec{r}(t)]\rangle_{\rm eff}$ and $|n[\vec{r}(t)]\rangle_s$, remains a slowly varying
function of time.

A straight forward calculation from the effective Hamiltonian (\ref{HIeff})
then gives the general expression for the geometric phase in an ARFP
\begin{eqnarray}
\gamma_{n}(t)&=&\int_{0}^{t}\nu_{nn}(t')dt'\nonumber\\
&=&-i\int_{0}^{t}\sum_{l}\left|\,_{\rm eff}\langle n[\vec{r}(t')|l\rangle_z\right|^{2}\times\nonumber\\
&&\,_s\langle
l[\vec{r}(t')]|\frac{d}{dt'}|l[\vec{r}(t')]\rangle_s dt'\nonumber\\
&&+\gamma^{(I)}_{n}(t),
\label{geometricphase}
\end{eqnarray}
and $\gamma^{(I)}_{n}(t)=-i\int_{0}^{t}dt'\,_{\rm eff}\langle
n[\vec{r}(t')|d/dt'|n[\vec{r}(t')\rangle_{\rm eff}$.
During the adiabatic motion in a given internal state, the
 time evolution of the coefficient $C_{n}(t)$
 takes the form
\begin{eqnarray}
C_{n}(t)=C_{n}(0)e^{-i\int_{0}^{t}\epsilon_{I}^{(n)}(t')dt'}e^{-i\gamma_{n}(t)}.
\end{eqnarray}

Equation (\ref{geometricphase}) is the central result of this
work. The geometric phase of an atom inside an ARFP is shown to
contain two parts. The second part, $\gamma^{(I)}_{n}(t)$, is
clearly due to the interaction term $\mu_{B}g_{F}\vec{F}\cdot\vec{B}^{{\rm
eff}}$ in $H_{\rm eff}^{(I)}$ (\ref{HIeff}),
with its value determined by the trajectory of the
direction for the ``effective B-field" $\vec{B}^{\rm eff}$. The
first part arises from the second term of
$H_{\rm eff}^{(I)}$ (\ref{HIeff}). It is determined by the
trajectories of both the static field $\vec{B}_{s}$ and the
effective B-field $\vec{B}^{\rm eff}$. The expression for
$\gamma_{n}$ in an ARFP is complicated because the internal quantum
state in an ARFP is assumed to be adiabatically kept in an
eigenstate of $\mu_{B}g_{F}\vec{F}\cdot\vec{B}^{\rm eff}$, rather
than an eigenstate of the total interaction Hamiltonian
$H^{(I)}_{\rm eff}(t)$.

In section IV, we will perform explicit calculations
for several examples of ARFP
proposed for various applications:
e.g., as atomic storage rings or atomic beam splitters.
Most often we find that only the first part of Eq. (\ref{geometricphase})
contributes a non-zero value to the geometric phase.

Before proceeding to the next section
for a quantal treatment of the geometric phase,
we find the time evolution of the
atomic spin state in the Schroedinger picture
\begin{eqnarray}
|\Psi(t)\rangle &=&\sum_{ml}C_{l}(0)\,_z\langle
m|l[\vec{r}(t)]\rangle_{\rm eff} \times\nonumber\\
&&e^{-i\int_{0}^{t}\epsilon_{I}^{l}(t')dt'}e^{-i\gamma_{l}(t)}e^{-im\omega
t}|m[\vec{r}(t)]\rangle_{s},
\label{precisestate}
\end{eqnarray}
obtained directly from $|\Psi(t)\rangle
=U^{\dagger}(t)|\Psi(t)\rangle_{I}$ after
the applications of the rotating
wave and adiabatic approximations.
When the atom is prepared initially in a specific adiabatic state
$|n[\vec{r}(t)]\rangle_{\rm eff}$ of the interaction picture, we arrive at
the simple case of $C_{l}(0)=\delta_{ln}$.

\section{A quantum mechanical treatment}

In the previous section, we provided the result for the geometric
phase $\gamma_{n}(t)$ in an ARFP based on a semi-classical approach,
where the atomic center of mass motion is described classically.
A clear physical picture exists in this case for the appearance of
the geometric phase in a certain parameter space.
The validity conditions for the rotating wave and the adiabatic
approximations as obtained above are all formulated
in terms of gauge independent forms. However, if the
influence of the geometric phase on the atomic spatial motion
is to be included, e.g., as in the
Aharonov-Bohm-type, phase shift, interference arrangement
in an atomic Sagnac interferometer discussed earlier \cite{peng-pra},
we would need an improved description where both the atomic spin and its
center of mass motion are treated quantum mechanically.

In a full quantum treatment of the atomic motion,
the quantum
state of an atom can be expressed as
$|\Phi(t)\rangle=\sum\phi_{l}(\vec{r},t)|l\rangle_z$, where
$\phi_{l}(\vec{r},t)$ is the atomic spatial wave function
for the internal state $|l\rangle_z$ of $F_{z}$.
The state $|\Phi(t)\rangle$ then satisfies the
Schroedinger equation governed by the Hamiltonian
\begin{eqnarray}
{\cal H}=\frac{{\vec
P}^2}{2M}+g_{F}\mu_{B}\vec{F}\cdot\vec{B}(\vec{r},t),
\end{eqnarray}
with $\vec{P}$ being the kinetic momentum and $M$ the atomic mass.

The rotating wave and adiabatic approximations can
be introduced now by defining the interaction picture
with the unitary transformation
\begin{eqnarray}
{\cal U}(t) &=&\left[\sum_{m=-F}^{F}|m\rangle_z\,_{\rm eff}\langle
m(\vec{r})|\right]\times\nonumber\\
&&\left[\sum_{n=-F}^{F}|n\rangle_z\,_s\langle
n(\vec{r})|e^{in\kappa\omega t}\right].\ \ \
\end{eqnarray}
The state in the interaction picture
$|\Phi(t)\rangle_{I}={\cal U}(t)|\Phi(t)\rangle$
now is governed by the Schroedinger equation
with the Hamiltonian
${\cal H}_{\rm eff}={\cal U}{\cal H}{\cal U}^{\dagger}$.
Under the rotating wave and adiabatic approximations,
we neglect transitions between states $|m\rangle_z$ and $|n\rangle_z$ $(m\neq n)$
as well as the rapidly oscillating terms.
We then obtain
\begin{eqnarray}
{\cal H}_{\rm eff}&\approx&\sum_{n}|n\rangle_z\,_z\langle n|{\cal H}_{\rm
eff}|n\rangle_z\,_z\langle n|\nonumber\\
&\approx& \sum_{n} H_{\rm ad}^{(n)}|n\rangle_z\,_z\langle n|,
\end{eqnarray}
where the adiabatic Hamiltonian
$H_{\rm ad}^{(n)}$ for the $n$-th adiabatic branch  is defined as
\begin{eqnarray}
H_{\rm ad}^{(n)}=\frac{\left(\vec{P}-\vec{A}_{n}\right)^{2}}{2M}
+\epsilon_{I}^{(n)}(\vec{r}),
\end{eqnarray}
with the effective gauge potential
\begin{eqnarray}
\vec{A}_{n}(\vec{r})&=&-i\sum_{l}\left|\,_{\rm eff}\langle
n(\vec{r})|l\rangle_z\right|^{2}\,_s\langle
l(\vec{r})|\nabla|l(\vec{r})\rangle_s\nonumber\\
&&-i\,_{\rm eff}\langle n(\vec{r})|\nabla|n(\vec{r})\rangle_{\rm
eff}. \label{An}
\end{eqnarray}

In this form, it is well known that the geometric phase
$\gamma_{n}$ can be expressed as the integral of the gauge
potential $\vec{A}_{n}$ along the spatial trajectory
for the atomic center of mass in an ARFP, i.e., one would expect generally that $\gamma_{n}=\int
\vec{A}_{n}\cdot d\vec{r}$. Similar to the result of the
semi-classical approach, the gauge potential
$\vec{A}_{n}(\vec{r})$ can be expressed as the sum of two parts.
The first part in Eq. (\ref{An}) is the weighted sum of the atomic
gauge potential $-i\,_s\langle
l(\vec{r})|\nabla|l(\vec{r})\rangle_s$ from the static field
$\vec{B}_{s}$, while the second term is the atomic gauge potential
from the ``effective B-field" $\vec{B}^{\rm eff}$.

A full quantum treatment for atomic
motion in an ARFP has been attempted earlier \cite{Rf-quantum}.
In fact, many of our formulations are identical to the results
of Ref. \cite{Rf-quantum}. For instance,
it is easy to show that the unitary transformations $U_{S}$, $U_{R}$,
and $U_{F}$ in \cite{Rf-quantum} are related directly to ours as
$U_{F}^{\dagger}U_{R}^{\dagger}U_{S}^{\dagger}={\cal U}$.
The only difference concerns the gauge potential $\vec{A}_{n}$
that was neglected in Ref. \cite{Rf-quantum}. Thus, they did not
give the expression for the gauge potential,
and the result for the geometric phase was not
obtained either \cite{Rf-quantum}.
Our study shows that the neglect of the adiabatic gauge potential
potentially can give rise to a final result, dependent
on the choice of the local phase factors for the internal eigenstate.

\section{Geometric phases in ARFP based applications}

In the above two sections, we obtain the expression for the
atomic geometric phase in an ARFP. This section is
devoted to the calculations of the geometric phases for
several proposed applications of ARFP, such as storage rings or beam
splitters for neutral atoms
\cite{Rf-ring,Rf-ring2,Rf-combine,Rf-classical,Rf-quantum}.

Before presenting our results for the more specific cases,
we provide some general discussions of the geometric phases
in several ARFP based storage rings. As was pointed out earlier, the
geometric phase $\gamma_{n}$ is given by the line integral of the
gauge potential $\vec{A}_{n}$ along the trajectory for the
atomic center of mass motion. For a
closed path in the storage ring at a fixed $\rho=\rho_c$ and $z=z_c$,
this can be further reduced to
\begin{eqnarray}
\gamma_{n}=q\int_{0}^{2\pi} A^{(\phi)}_{n}(\rho,\phi,z)\rho d\phi,
\label{ringphase}
\end{eqnarray}
where the integer $q$ is the
winding number of the path and
$A^{(\phi)}_{n}$ is the component of $\vec{A}_{n}$ along the
azimuthal direction $\hat{e}_{\phi}$ of the familiar cylindrical
coordinate system $(\rho,\phi,z)$. Without loss of
generality, we take $q=1$ in this paper.
For the storage rings proposed in Refs.
\cite{Rf-ring,Rf-ring2,Rf-combine,Rf-classical,Rf-quantum},
the gauge potentials $A^{(\phi)}_{n}(\rho,\phi,z)$ are
actually independent of the angle $\phi$.
Therefore, the geometric phase is simply given by
\begin{eqnarray}
\gamma_{n}^{(c)}=2\pi \rho_{c}A^{(\phi)}_{n}(\rho_{c},z_{c}),
\end{eqnarray}
given out in explicit forms for different
storage ring schemes
\cite{Rf-classical,Rf-quantum,Rf-combine,Rf-ring,Rf-ring2}.

In reality, because of thermal motion or when the
atomic transverse motional state is considered,
the center of mass for an atom
can deviate from $(\rho_c,z_{c})$ even for a closed trajectory.
This uncertainty in the exact shape of the closed trajectory gives
rise to a fluctuating geometric phase and is usually difficult
to study. Assuming a simple closed path at
fixed $\rho$ and $z$, we have found previously that
the subsequently fluctuations could
decrease the visibility of the interference pattern
\cite{peng-pra}. Quantum mechanically, such destructive interference
can be explained as resulting from
entanglement between the freedoms for
$\phi$ and $(\rho,z)$ because of the
dependence of the gauge potential $A_{n}^{\phi}$ on
$\rho$ and $z$. Therefore, it is important to investigate this
dependence near the trap center.

For simplicity, our discussions below will focus on the closed
loops where $\rho$ and $z$ are $\phi$-independent constants.
In this case, the geometric phase can be expressed as
$\gamma_{n}(\rho,z)=2\pi \rho A^{(\phi)}_{n}(\rho,z)$.
We will show numerically the distributions for
$\gamma_{n}(\rho,z)$ obtained this way near the central region
of $(\rho_c,z_{c})$. If needed, a more rigorous approach can be
developed to investigate the fluctuations of the resulting geometric
phase from the gauge potential $A^{(\phi)}_{n}(\rho,z)$.

\subsection{The storage ring proposals of Refs. \cite{Rf-ring,Rf-ring2,Rf-combine}}

This subsection is devoted to a detailed calculation of the
geometric phases for the ARFP storage ring proposals of Refs.
\cite{Rf-ring,Rf-ring2,Rf-combine}. We will derive the analytical
expressions for the azimuthal component $A^{(\phi)}_{n}$ of the
gauge potential that arises in both cases from cylindrically
symmetric static B-field and rf fields. Because of the cylindrical
symmetry, the angle $\beta_{s}(\rho,z)$ between the local static
B-field and the $z$-axis is required to be analytical in the
region near the storage ring. Therefore, the eigenstate
$|n(\vec{r})\rangle_{s}$ of $\vec{F}\cdot\vec{B}_{s}$ can be
chosen as
\begin{eqnarray}
|n(\vec{r})\rangle_{s}=\exp\{-i[\vec{F}\cdot\hat{e}_{\phi}\beta_{s}(\rho,z)+n\phi]\}|n\rangle_z.
\label{ns}
\end{eqnarray}
Consequently, $\vec{B}^{\,\rm eff}(\vec{r})$ is also cylindrically
symmetric, which leads to the eigenstate $|n(\vec{r})\rangle_{\rm eff}$
of $\vec{F}\cdot\vec{B}^{\,\rm eff}$ as
\begin{eqnarray}
|n(\vec{r})\rangle_{\rm eff}=\exp\{-i[\vec{F}\cdot\hat{n}^{\,\rm
eff}_{\perp}(\vec{r})\beta_{\rm eff}(\rho,z)+n\phi]\}|n\rangle_z,
\label{neff}
\end{eqnarray}
with the unit vector $\hat{n}^{\,\rm eff}_{\perp}(\vec{r})$
in the $x$-$y$ plane orthogonal to $\vec{B}^{\,\rm eff}(\vec{r})$
 and $\beta_{\,\rm eff}(\rho,z)$ denoting the angle between
$\vec{B}^{\,\rm eff}(\vec{r})$ and the $z$-axis.
We note that the unit vector field $\hat{n}^{\,\rm eff}_{\perp}(\vec{r})$
also possesses cylindrical symmetry, i.e., remains invariant under
rotation around the $z$-axis. The expressions of (\ref{ns}) and (\ref{neff})
allow us to obtain
the simple expression of the gauge potential
\begin{eqnarray}
A^{(\phi)}_{n}(\rho,z)=-\frac{n}{\rho}\cos\beta_{\rm
eff}(\rho,z)\cos\beta_{s}(\rho,z),
\label{Aring}
\end{eqnarray}
after straightforward calculations.

In the scheme of Ref. \cite{Rf-ring},
the static B-field is a ``ring-shaped quadrupole field" that
vanishes along a circle of a radius $\rho_0$ in the $x$-$y$ plane.
Near $\rho=\rho_0$, the B-field is given approximately by
\begin{eqnarray}
\vec{B}_{s}(\vec{r})=B'(\rho-\rho_{0})\hat{e}_{\rho}-B'z\hat{e}_{z},
\label{ring1bs}
\end{eqnarray}
like a quadrupole field, while the rf-field takes
a complicated form
\begin{eqnarray}
\vec{B}_{o}(\vec{r},t)&=&\left(\frac{a}{\sqrt{2}}\cos(\omega
t)+\frac{b}{\sqrt{2}}\cos(\omega
t+\varphi)\right)\hat{e}_{\rho}\nonumber\\
& &+\left(-\frac{a}{\sqrt{2}}\sin(\omega
t)+\frac{b}{\sqrt{2}}\sin(\omega
t+\varphi)\right)\hat{e}_{z}, \ \ \ \ \ \
\label{ring1rf}
\end{eqnarray}
with constants $a$ and $b$ independent of $\vec{r}$.

\begin{figure}
\includegraphics[height=2.5in]{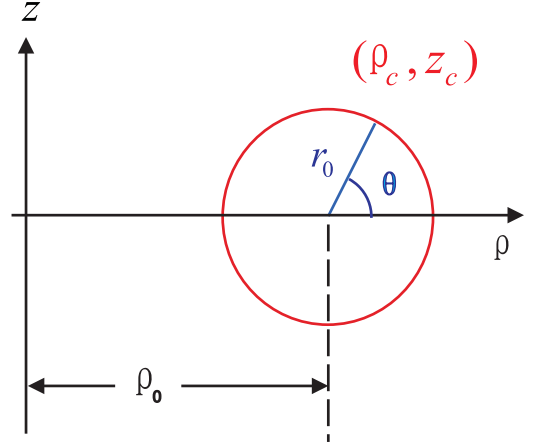}
\caption{(Color online) A cross-sectional view for the storage
ring of Ref. \cite{Rf-ring}. The static field is zero in the ring
at the fixed radius $\rho_{0}$. The addition of rf-fields creates
an ARFP centered at a ring through $(\rho_{c},z_{c})$. The distance
from the trap center to the ring with radius $\rho_{0}$ in the
plane $z=0$ is $r_{0}$.} \label{fig1}
\end{figure}

From the expression of (\ref{ns}) for the eigenstate
$|n(\vec{r})\rangle_{s}$, the ``effective B-field" $\vec{B}^{\,\rm
eff}$ becomes
\begin{eqnarray}
\vec{B}^{\,\rm
eff}(\vec{r})&=&B'[\sqrt{(\rho-\rho_{0})^{2}+z^{2}}-r_{0}]\hat{e}_{z}\nonumber\\
&&-\left(\frac{b}{\sqrt{2}}\cos(\theta+\varphi)+\frac{a}{\sqrt{2}}\cos\theta\right)\hat{e}_{\rho}\nonumber\\
&&+\left(\frac{b}{\sqrt{2}}\sin(\theta+\varphi)-\frac{a}{\sqrt{2}}\sin\theta\right)\hat{e}_{\phi},
\end{eqnarray}
where $r_{0}$ and $\theta$ are given by
\begin{eqnarray}
r_{0}&=&\frac{\omega}{|\mu_{B}g_{F}B'|}\,,\nonumber\\
\cos\theta(\rho,z)&=&\frac{\rho-\rho_{0}}{\sqrt{(\rho-\rho_{0})^{2}+z^{2}}}\,, \nonumber\\
\sin\theta(\rho,z)&=&\frac{z}{\sqrt{(\rho-\rho_{0})^{2}+z^{2}}}\,.\label{theta}
\end{eqnarray}

In an ARFP, as discussed here, the trap center at
$(\rho_{c},z_{c})$ is determined by minimizing both the
$z$-component and the transverse component of $\vec{B}^{\,\rm eff}$.
Without loss of generality, we will assume $a,b>0$. Then,
$(\rho_{c},z_{c})$ is found to satisfy
\begin{eqnarray}
\theta(\rho_{c},z_{c})&=&-\varphi/2,\nonumber\\
\sqrt{(\rho_{c}-\rho_{0})^{2}+z_{c}^{2}}&=&r_{0},
\end{eqnarray}
i.e., the trap center lies on the surface of
the ``resonance toroid" at $\rho=\rho_0$ with a radius $r_{0}$
as shown in Fig. \ref{fig1}.
The relative angle of the trap center with respect to
the center of the toroid cross-section is given by $-\varphi/2$.
On this ``resonance toroid," the
rf-field is resonant with the static field, i.e.,
$B^{\,\rm eff}_{z}$ vanishes. As a result,
the ``effective B-field" lies again in the $x$-$y$ plane on the ``resonance toroid,"
which gives $\cos\beta_{\rm eff}(\rho_{c},z_{c})=0$ and leads to
the result $A^{(\phi)}_{n}(\rho_{c},z_{c})=\gamma_{n}=0$ as shown
in the trap center for the storage ring considered before in Ref.
\cite{Rf-ring}.

From the expression (\ref{ring1bs}) of the static field and
the definition of the angle $\theta(\rho,z)$, we find
a simple relationship $\beta_{s}(\rho,z)=\pi/2+\theta(\rho,z)$,
with which the gauge potential $A^{(\phi)}_{n}(\rho,\phi)$ in
(\ref{Aring}) can be further simplified as
\begin{eqnarray}
A^{(\phi)}_{n}(\rho,z)&=&\frac{n}{\rho}\cos\beta_{\rm
eff}(\rho,z)\sin\theta(\rho,z)\nonumber\\
&\approx&\frac{n}{\rho}\cos\beta_{\rm
eff}(\rho,z)\sin\theta(\rho_{c},z_{c}),
\end{eqnarray}
near the trap center. Thus, the spatial fluctuation of the gauge potential
$A^{(\phi)}_{n}(\rho,z)$ in the region around the trap center is
closely related to the angle $\theta(\rho_{c},z_{c})$ of the
trap center, or the parameter $\varphi$ of the oscillating field
$\vec{B}_{o}$. When $\varphi=0$, the atom is trapped in the region
with $\theta\approx 0$ or $\pi$, where the fluctuation of
$A^{(\phi)}_{n}(\rho,z)$ is suppressed significantly due to
the small value of $\sin\theta$. On the other
hand, if the angle $\varphi$ is set to $\pi$ with the trap
center located in the region with $\theta\approx
\pm\pi/2$, the fluctuation of the gauge potential becomes amplified.

\begin{figure}
\includegraphics[height=2.35in]{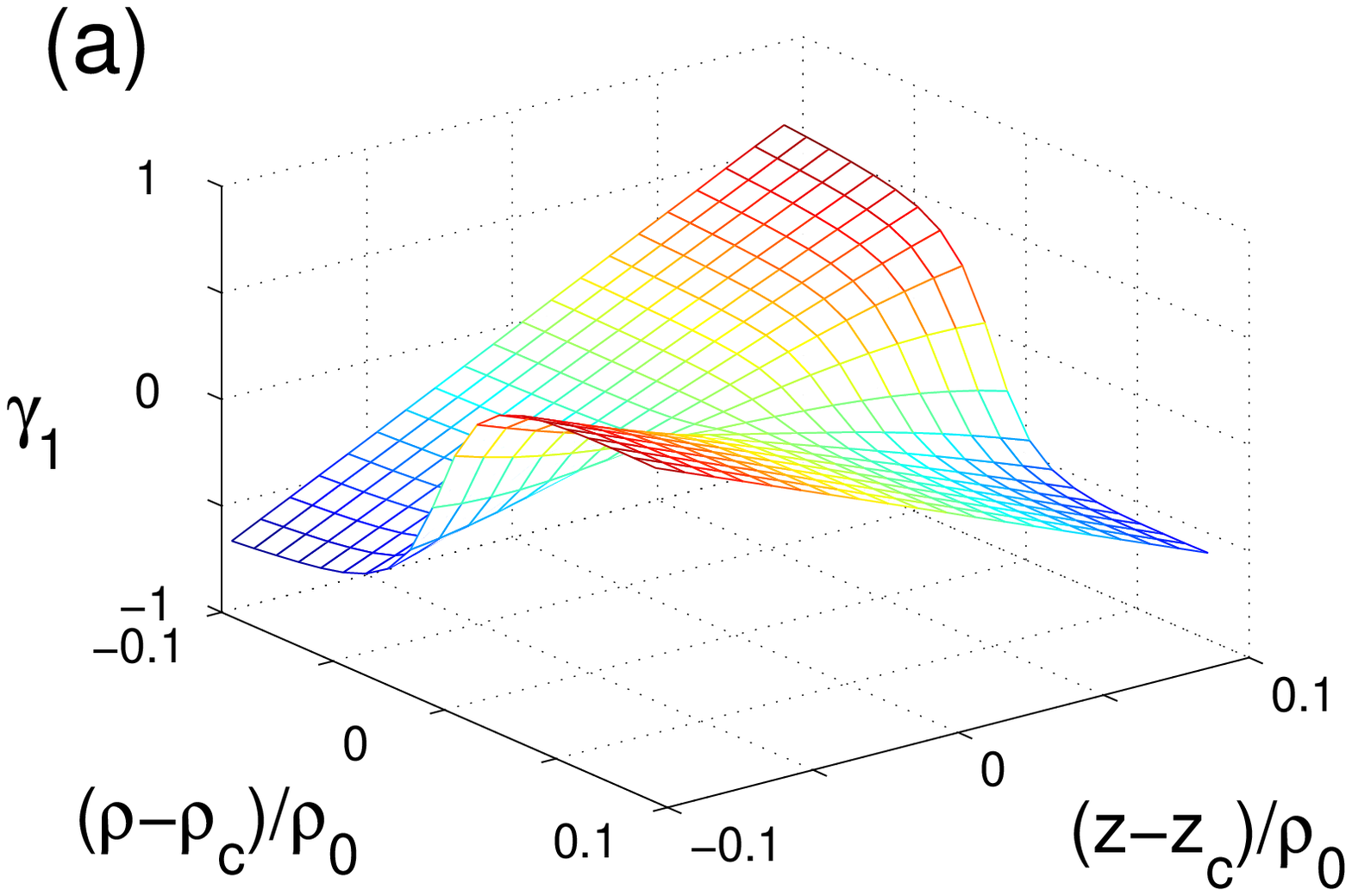}
\includegraphics[height=2.35in]{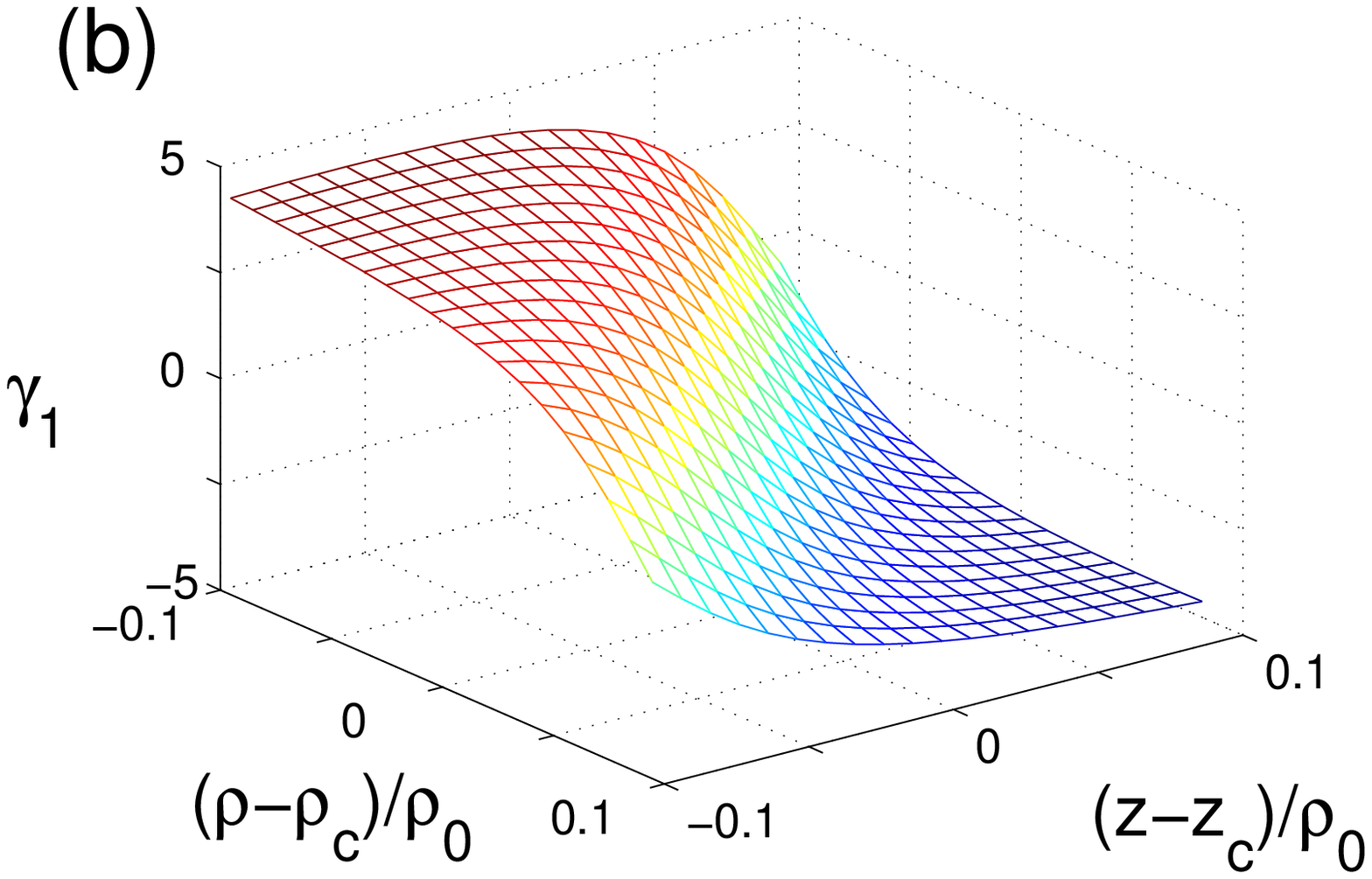}
\includegraphics[height=2.35in]{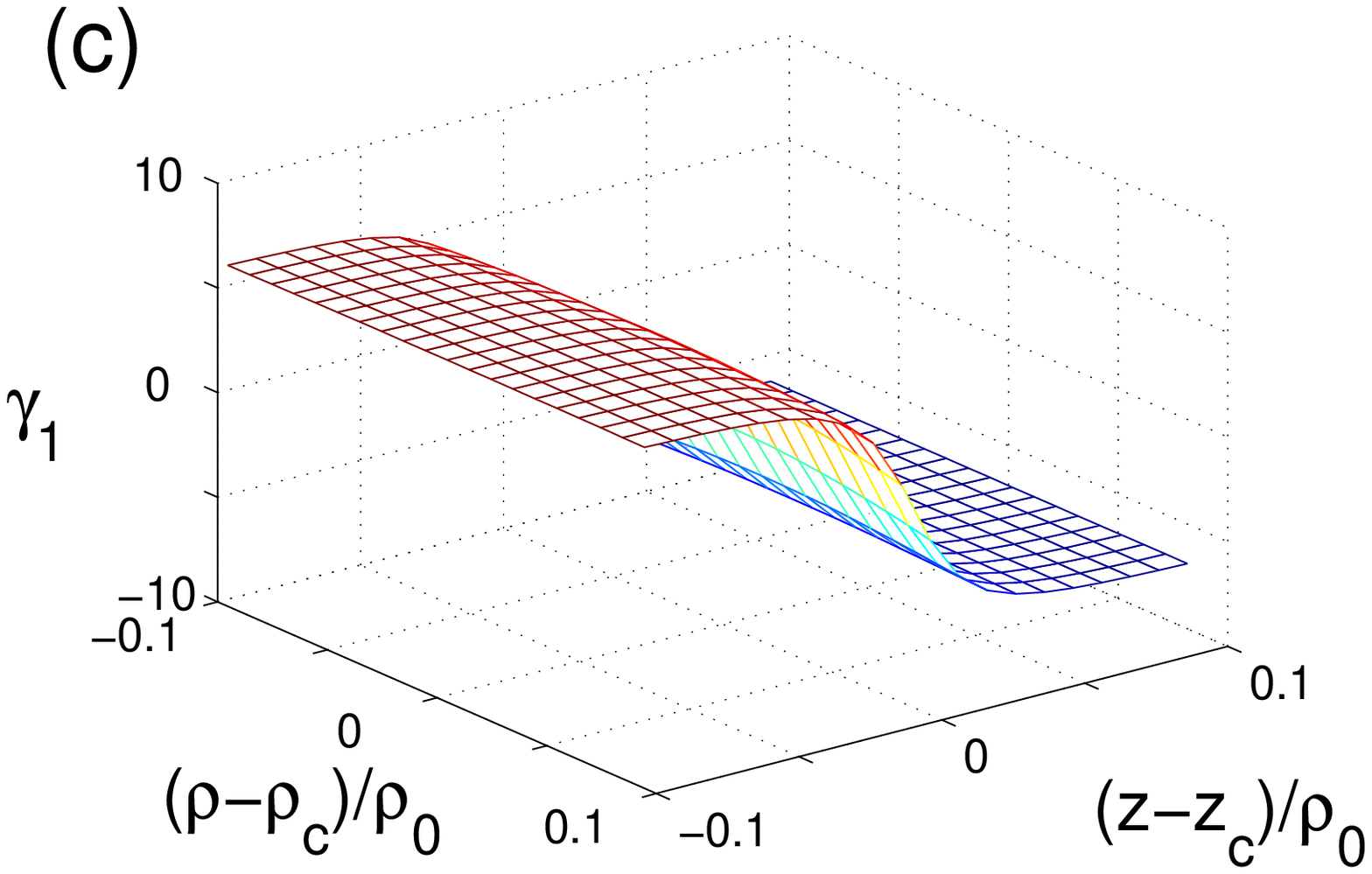}
\caption{(Color online) The distribution of the geometric
phase $\gamma_{1}$ near the trap center $(\rho_{c},z_{c})$
of the storage ring proposed in Ref. \cite{Rf-ring}
at (a) $\varphi=0$, (b) $\varphi=\pi/2$,
and (c) $\varphi=\pi$, clearly displaying the
$\sin(\varphi/2)$ dependence.}
\label{fig2}
\end{figure}

In Fig. \ref{fig2}, we illustrate numerical results for the
distribution of the geometric phase $\gamma_{1}(\rho,z)=2\pi\rho
A_{1}^{\phi}(\rho,z)$ in the region near the trap center at
$\varphi=0,\pi/2,\pi$. We see clearly decreased fluctuations
of $\gamma_{1}$ when the absolute value of
$\sin\theta(\rho_{c},z_{c})=-\sin(\varphi/2)$ is decreased.

Next we turn to the storage ring of Ref. \cite{Rf-ring2}
constructed from a quadrupole static B-field
$\vec{B}_{s}(\vec{r})=B'(x,y,-2z) $
and an $\vec{r}$-independent rf field $\vec{B}_{o}=B_{\rm
rf}\cos(\omega t)\hat{e}_{z}$ along the $z$ direction.
The resulting ARFP provides a 2D ring shaped trap in the $x$-$y$ plane.
In addition, a 1D optical potential along the $z$ direction is
employed to confine atoms in the transverse plane at
$z=0$ \cite{Rf-ring2}.
The ``effective B-field" takes the form
\begin{eqnarray}
\vec{B}^{\,\rm
eff}(\vec{r})=B'(\rho-\rho_{0})\hat{e}_{z}-\frac{1}{2}B_{\rm
rf}\hat{e}_{\rho},
\end{eqnarray}
in the plane at $z=0$, with
$\rho_{0}=\omega/|\mu_{B}g_{F}B'|$. Because the strength
of $\vec{B}^{\,\rm eff}$ is near minimum at the ring $\rho=\rho_{0}$,
the trap center for this storage ring is
located at $\rho_{c}=\rho_{0}$ and $z_{c}=0$.
At the trap center, the ``effective B-field" is along the
direction of $\hat{e}_{\rho}$.
Thus, according to Eq. (\ref{Aring}), the geometric phase
$\gamma_{n}^{(c)}$ at the trap center again vanishes.

\begin{figure}
\includegraphics[height=2.35in]{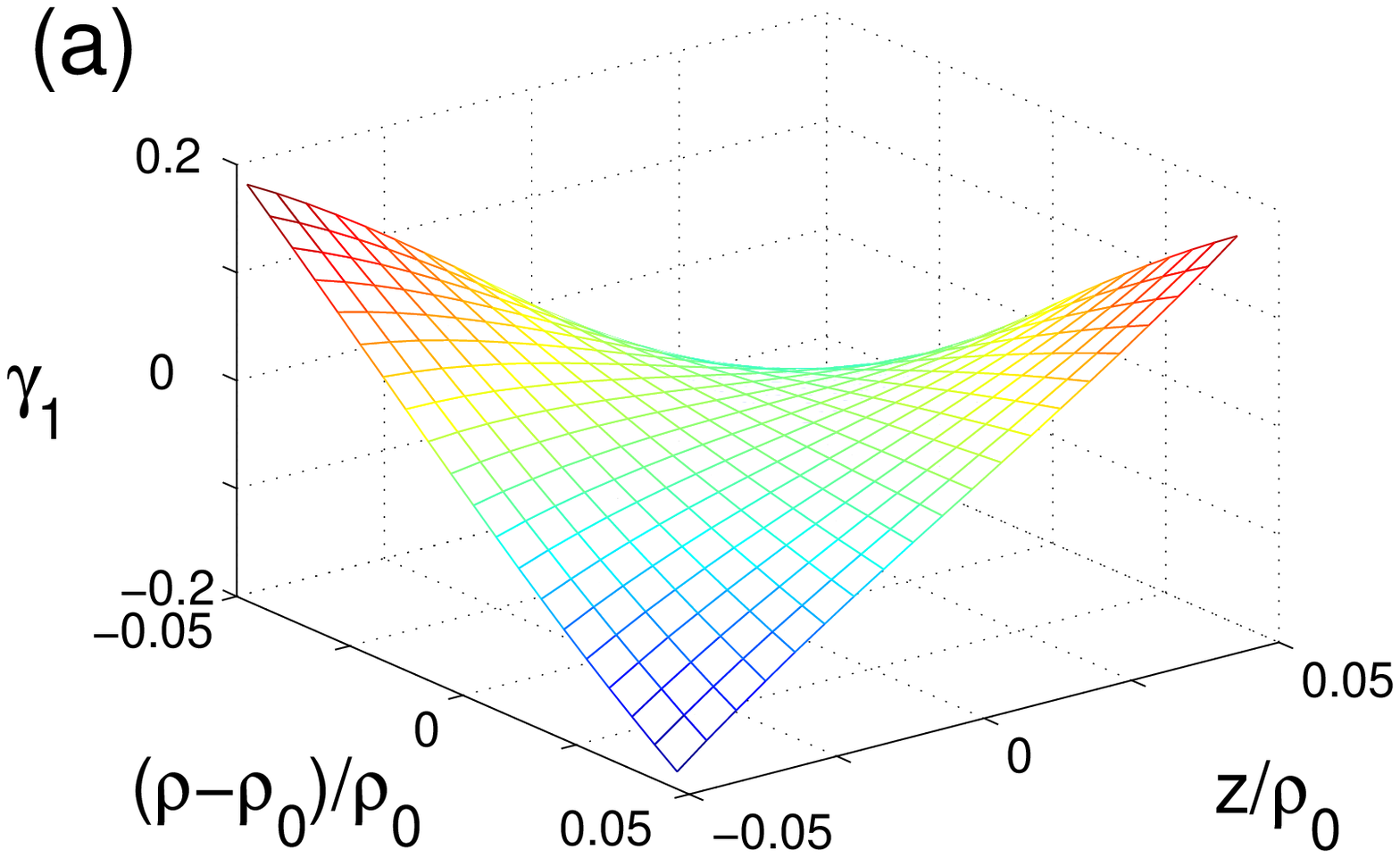}
\includegraphics[height=2.35in]{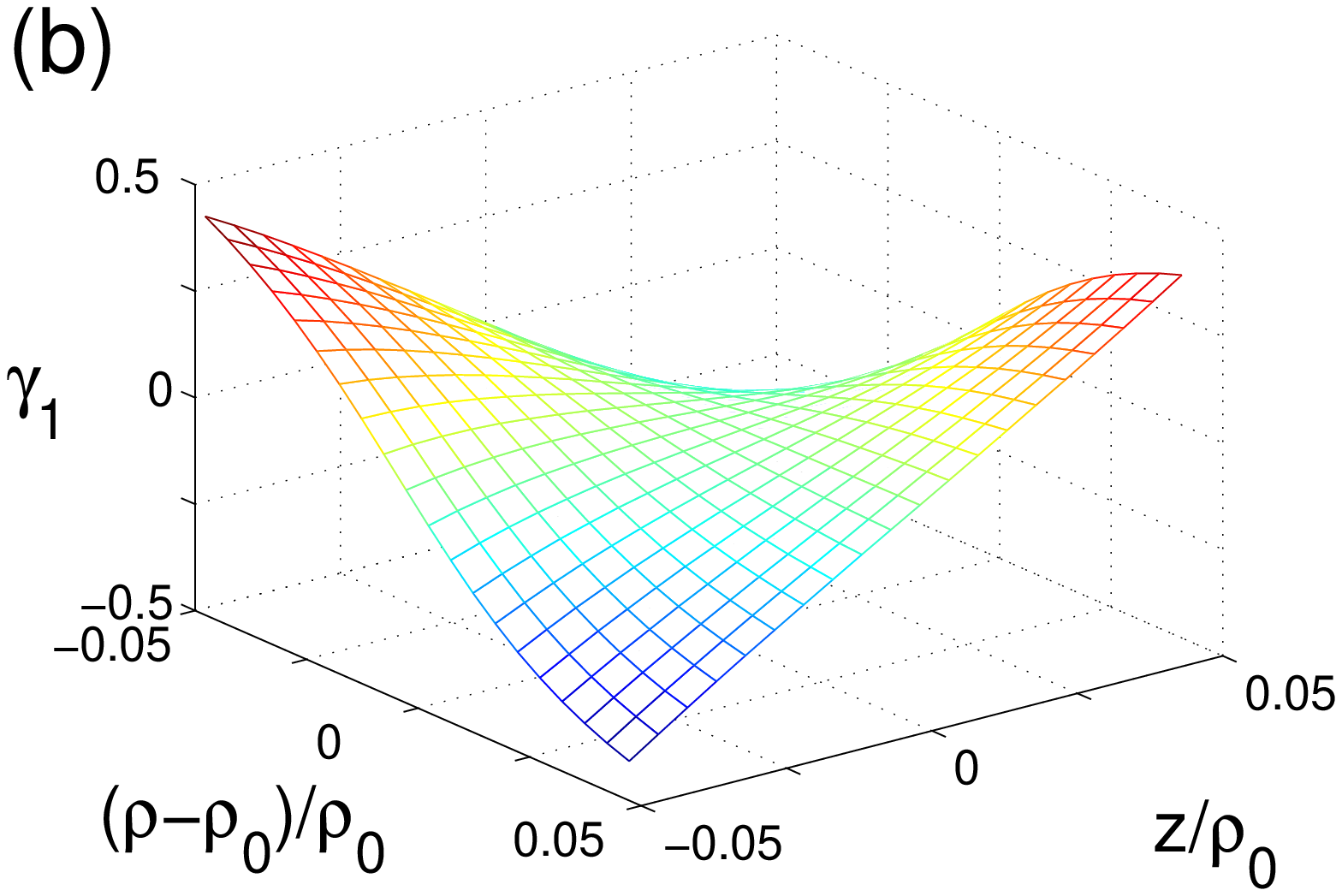}
\caption{(Color online) The geometric phase $\gamma_{1}$ for the
storage ring of Ref. \cite{Rf-ring2} with (a) $B_{\rm rf}=$
$0.15B'\rho_{0}$ and (b) $B_{\rm rf}=0.05B'\rho_{0}$.}
\label{fig3}
\end{figure}

In Fig. \ref{fig3}, we show the distribution of the geometric
phase $\gamma_{1}$ in the region near the trap center for $B_{\rm
rf}=0.05|B'|\rho_{0}$ and $B_{\rm rf}=0.15|B'|\rho_{0}$. We see
that the fluctuation is relatively small when the strength of the
rf-field is large. This can be explained by
Eq. (\ref{Aring}), which shows that $A_{n}^{(\phi)}$ is
proportional to $\cos\beta_{\rm eff}$ and can be approximated as
$2B^{\,\rm eff}_{z}/B_{\rm rf}$ near the trap center. When $B_{\rm
rf}$ is large, the gauge potential becomes a relatively slow
varying function of $\rho$ and $z$. In this case, the presence
of a 1D optical potential allows for the possibility of tuning the
trap center position to a nonzero value of $z$, with the storage
ring remaining in the $x$-$y$ plane. Then $\cos\beta_{s}$ is assumed to
a nonzero value, leading to increased fluctuations for the
geometric phase.

Finally, we discuss the geometric phase
in the ``time averaged" ARFP storage ring proposed in
Ref. \cite{Rf-combine}. Unlike previously
considered ARFP based storage rings, the time dependence
now exists in both the ``static B-field" and the frequency
of the rf field given by
\begin{eqnarray}
\vec{B}_{s}(\vec{r},t)&=&B'\rho\hat{e}_{\rho}-2B'z\hat{e}_{z}+B_{m}\sin(\omega_{m}t)\hat{e}_{z},\nonumber\\
\vec{B}_{o}(t)&=&B_{\rm rf}\sin[\omega(t)t]\hat{e}_{z},\nonumber\\
\omega(t)&=&\omega_{0}\sqrt{1+(B_{m}/B'\rho_{0})^{2}\sin^{2}(\omega_{m}t)}\,.
\end{eqnarray}
The frequency $\omega_{m}$ is assumed to be much smaller than
$\omega_{0}$ but much larger than the trap frequency.
The radius $\rho_{0}$ is now defined as $\rho_{0}=\omega_{0}/|\mu_{B}g_{F}B'|$,
and the ``effective B-field" takes the form
\begin{eqnarray}
\vec{B}^{\,\rm
eff}(\vec{r},t)=\Delta(\vec{r},t)\hat{e}_{z}-\frac{B'\rho}{|2\vec{B}_{s}(\vec{r},t)|}B_{\rm
rf}\hat{e}_{\phi}\,.
\end{eqnarray}

\begin{figure}
\includegraphics[height=2.35in]{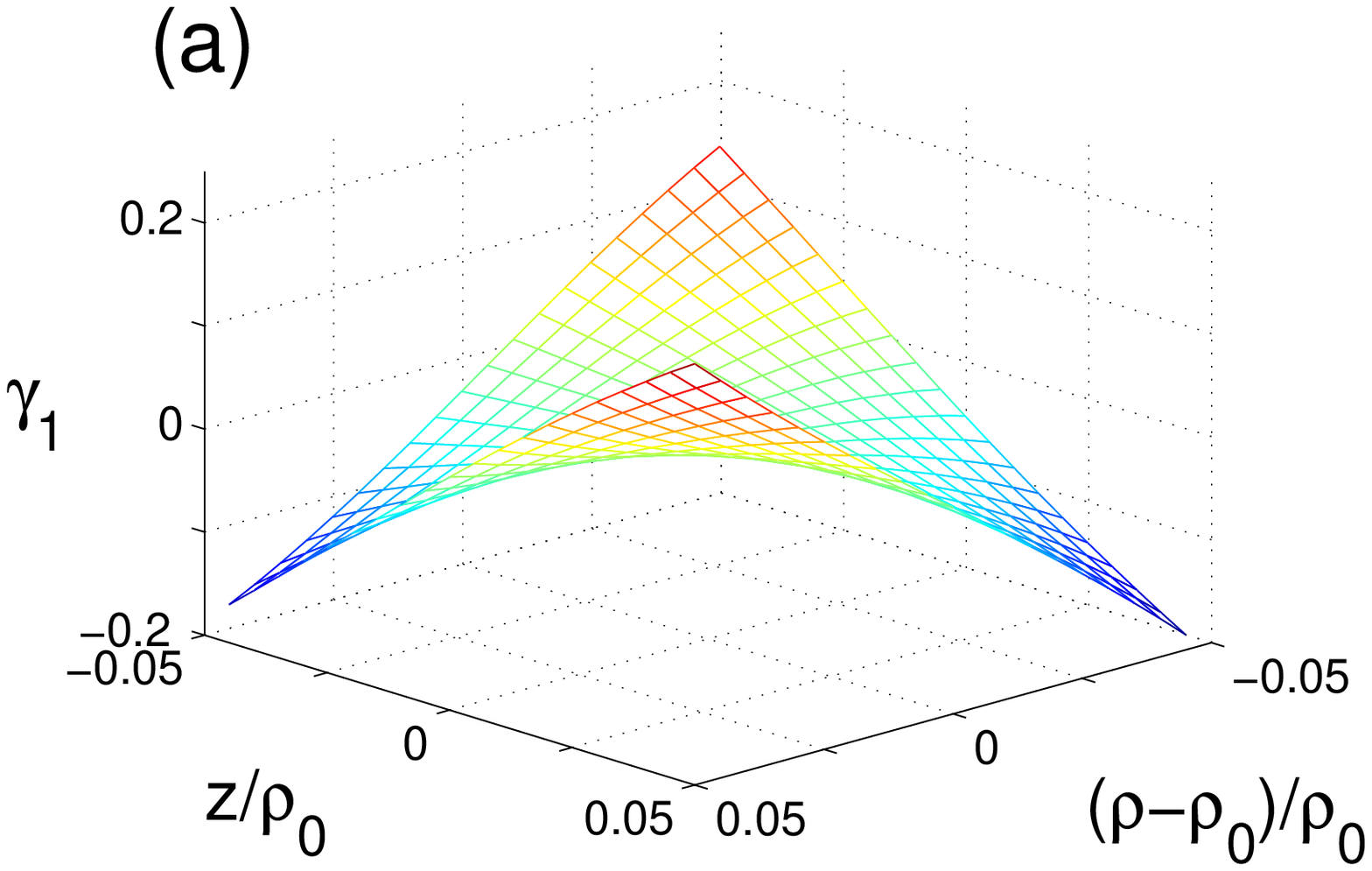}
\includegraphics[height=2.35in]{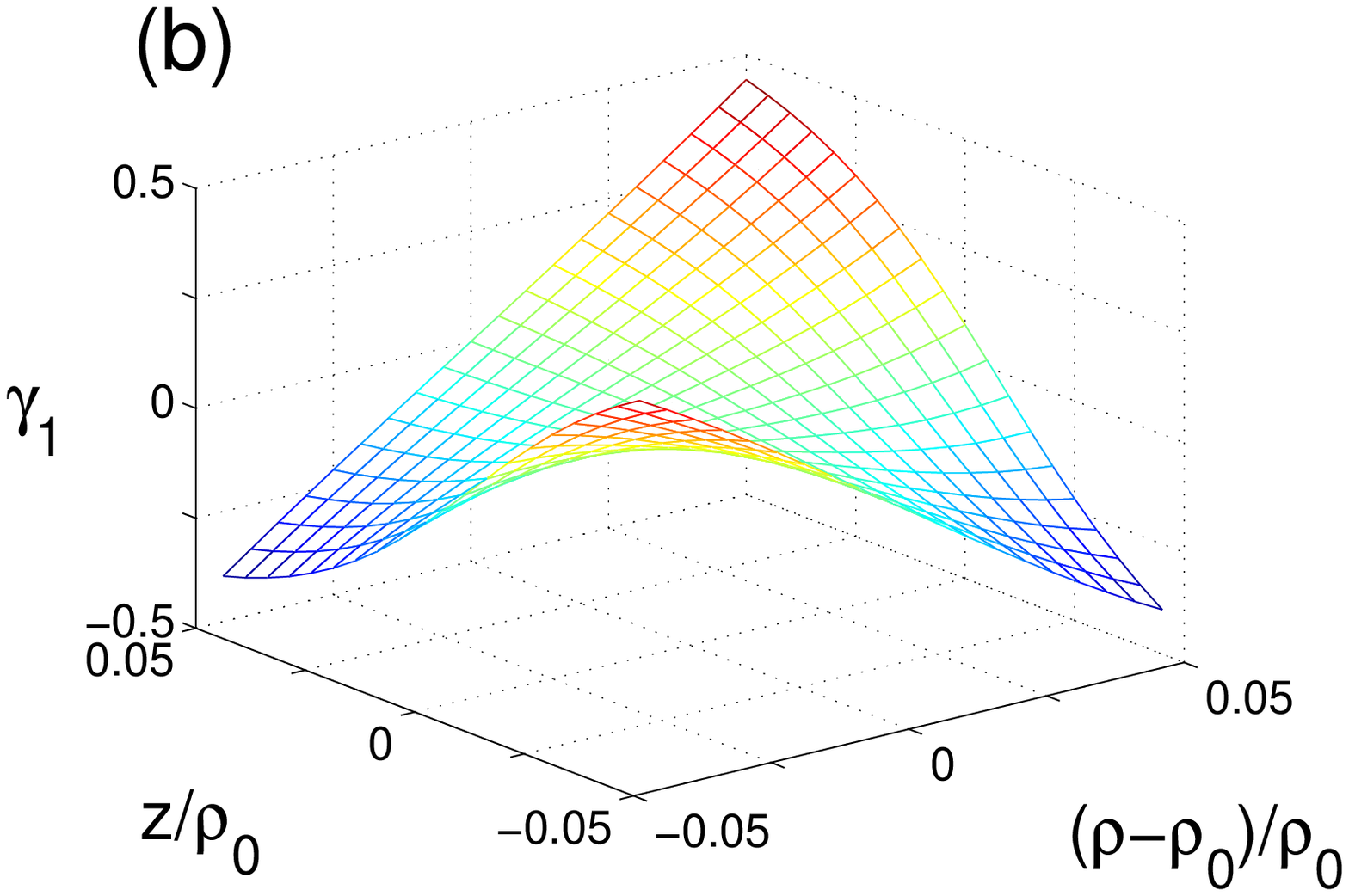}
\caption{(Color online) The geometric phase $\gamma_{1}$ for the
storage ring of Ref. \cite{Rf-combine} at (a) $B_{\rm
rf}=0.3B'\rho_{0}$ and (b) $B_{\rm rf}=0.1B'\rho_{0}$.
$B_{m}=0.05B'\rho_{0}$. }
\label{fig4}
\end{figure}

The operating principle for the time averaged storage ring of Ref.
\cite{Rf-combine} is similar to the well-known TOP \cite{TOP} and
TORT traps \cite{TORT,Kurn1}. The effective trap potential
experienced by the atom is proportional to the time averaged value
of the ``effective B-field"
$\int_{0}^{2\pi/\omega_{m}}|\vec{B}^{\,\rm eff}(\vec{r},t)|dt$.
When $B_{\rm rf}$ and $B_{m}$ are much smaller than $B'\rho_{0}$,
the center of the storage ring is located approximately at
$\rho_{c}=\rho_{0},z_{c}=0$. Using the earlier result
\cite{peng-pra}, we find that in the time averaged storage ring, the effective
gauge potential $\tilde{A}_{n}^{(\phi)}(\rho,z)$ is reduced simply
to the time averaged instantaneous gauge potential
\begin{eqnarray}
\tilde{A}_{n}^{(\phi)}(\rho,z)=\frac{\omega_{m}}{2\pi}\int_{0}^{2\pi/\omega_{m}}
A_{n}^{(\phi)}(\rho,z,t)dt,
\end{eqnarray}
with $A_{n}^{(\phi)}(\rho,z,t)$ given in (\ref{Aring}). The
geometric phase then is given approximately by
$\gamma_{n}(\rho,z)=2\pi\tilde{A}_{n}^{(\phi)}(\rho,z)$. In this
case, we find that the geometric phase always vanishes at the trap
center $(\rho_{c},z_{c})$. Figure \ref{fig4} illustrates the
distribution of the geometric phase in the region near the trap
center for two different values of the rf-field amplitude $B_{\rm
rf}$. Similar to the storage ring of Ref. \cite{Rf-ring2}, the
fluctuation of the geometric phase is suppressed in this case for
large $B_{\rm rf}$.

\subsection{The storage ring proposals of Refs. \cite{Rf-classical,Rf-quantum}}

Next we consider the ARFP based storage ring proposed in Refs.
\cite{Rf-classical,Rf-quantum}. In this case,
the static B-field is that of a Ioffe-Pritchard trap
on an atom chip. In the Cartesian coordinate $(x,y,z)$,
it takes the form
\begin{eqnarray}
\vec{B}_{s}=B'x\hat{e}_{x}-B'y\hat{e}_{y}+B'L\hat{e}_{z},
\label{IPT}
\end{eqnarray}
where $B'$ is the B-field gradient and the bias field along
the $z$-direction is denoted as $B'L$. The amplitudes
$\vec{B}^{(a)}_{\rm rf}$ and $\vec{B}^{(b)}_{\rm rf}(z)$ of the rf
field are $ \vec{B}^{(a)}_{\rm rf}=[B_{\rm
rf}(z)/\sqrt{2}]\hat{e}_{x}$ and $ \vec{B}^{(b)}_{\rm rf}=[B_{\rm
rf}(z)/\sqrt{2}]\hat{e}_{y}$ with
\begin{eqnarray}
B_{\rm rf}(z)=B_{\rm rf}^{(0)}+B''z^{2}.
\end{eqnarray}

In the schemes of Ref. \cite{Rf-classical,Rf-quantum} considered
earlier, the phase $\eta$ of the rf field is assumed to be
$\kappa\pi/2$. The $x$- and $y$-components of the ``effective
B-field" $\vec{B}^{\,\rm eff}(\vec{r})$ then become
\begin{eqnarray}
B^{\,\rm eff}_{x}(\vec{r})&=&\frac{B_{\rm rf}(z)}{2\sqrt{2}}(1+\cos\beta_{s}(\rho,z)),\nonumber \\
B^{\,\rm eff}_{y}(\vec{r})&=&0, \label{BIx}
\end{eqnarray}
according to Eq. (\ref{BI}). Then the strength of the ``effective
B-field" $\vec{B}^{\,\rm eff}$ has its minimum along a circle with
a non-zero radius $\rho_c$, provided a positive detuning $\Delta$
exists at the origin $(0,0,0)$ \cite{Rf-classical,Rf-quantum}. The
``effective B-field" $\vec{B}^{\,\rm eff}$ is easily shown to lie
in the $x$-$z$ plane along the trap bottom mapped out by the
atomic center of mass motion. This gives rise to a vanishing
$\gamma_{F}^{(I)}$. With a proper choice for the local phase
of $|n(\vec{r})\rangle_{\rm eff}$, the gauge potential
$A^{(\phi)}_{n}$ takes the form
\begin{eqnarray}
A^{(\phi)}_{n}(\rho,z)=\frac{n}{\rho}\cos\beta_{\rm
eff}(\rho,z)\left(1-\cos\beta_{s}(\rho,z)\right).
\end{eqnarray}

\begin{figure}
\includegraphics[height=2.35in]{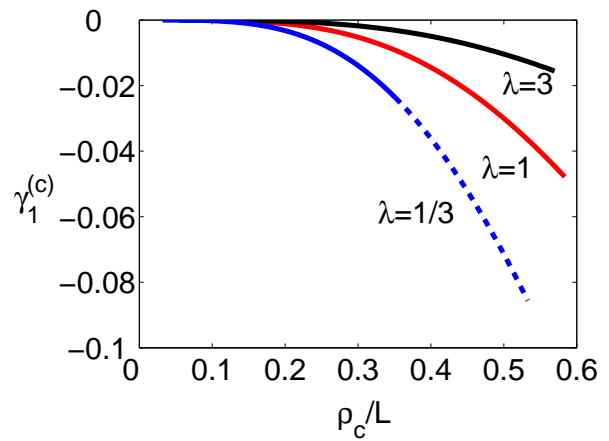}
\caption{(Color online) The geometric phase $\gamma_{1}$ is
plotted against the radius $\rho_{c}$ for the ARFP storage ring
of Ref. \cite{Rf-classical,Rf-quantum} with $\eta=\kappa\pi/2$ at
$\lambda=3$, $\lambda=1$, and $\lambda=1/3$. To assure the
validity of the rotating wave approximation, in the solid lines,
the maximal value of $\Delta[\vec{r}=0]/|g_{F}\mu_{B}B'\rho_{0}|$
or $B_{\rm rf}/(\sqrt{2}B'\rho_{0})$ are restricted to be smaller
than $0.15$. The extending dashed line is beyond the
 rotating wave approximation for $\lambda=1/3$
and $B_{\rm rf}/(\sqrt{2}B'\rho_{0})\in[0.15,0.3]$.}
\label{fig5}
\end{figure}

Figure \ref{fig5} displays the geometric phase along a closed path
for a spin-$1$ atom as a function of $\rho_c$ for the ARFP storage
ring proposed in Refs. \cite{Rf-classical,Rf-quantum}. The
parameter $\lambda$ is defined as
\begin{eqnarray}
\lambda=\sqrt{2}\frac{\Delta[\vec{r}=0]}{|g_{F}\mu_{B}B_{\rm
rf}^{(0)}|}\ .
\end{eqnarray}

To assure the validity of the rotating wave approximation, we find
that the maximal values of $\Delta[\vec{r}=0]/(|g_{F}|\mu_{B})$ and
$B_{\rm rf}^{(0)}/\sqrt{2}$ must be restricted to the region of
$\lambda\in [0,0.15]$.

As shown in Fig. \ref{fig1}, the geometric phase remains much
smaller than $2\pi$ in this situation. This fact can be
appreciated easily if we look at the distribution of the ``effective
B-field" ${\vec B}^{\,\rm eff}$. According to Eq. (\ref{BIx}), the
component $B_{x}^{\,\rm eff}$ has a nonzero minimal value $B_{\rm
rf}/(2\sqrt{2})$, while $|B^{\rm eff}_{z}|$ can become arbitrarily small,
although not necessarily zero in general. Therefore, at the trap
center where $|{\vec B}^{\,\rm eff}|$ is a minimum, the value of
$\cos\beta_{\rm eff}=B^{\,\rm eff}_{z}/|{\vec B}^{\,\rm eff}|$ can
become very small, leading to small geometric phases. Yet, despite
the relatively small geometric phase found here, our result
remains important because it could represent a systematic error if
not properly included in a Sagnac interference experiment.

\begin{figure}
\includegraphics[height=2.35in]{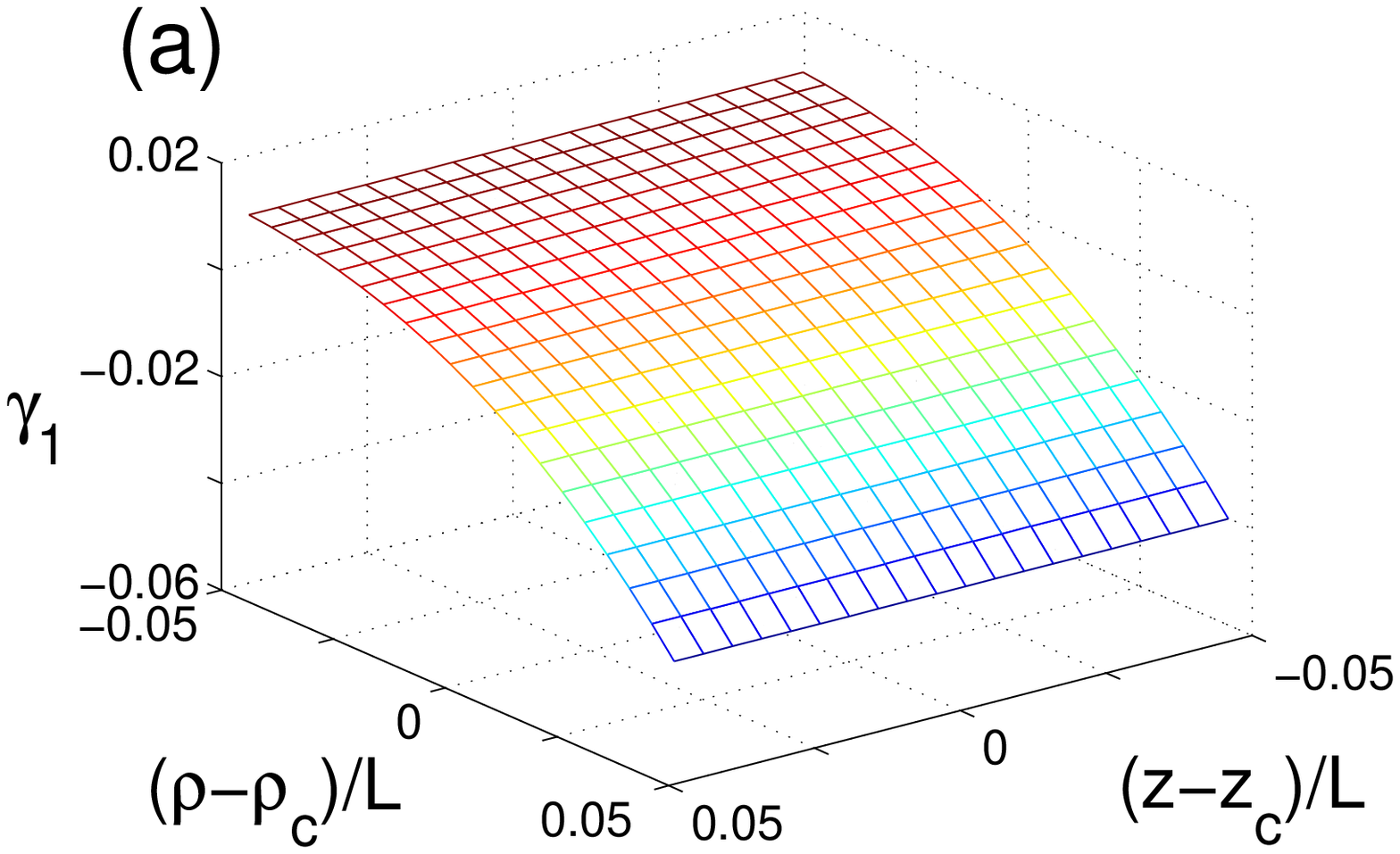}
\includegraphics[height=2.35in]{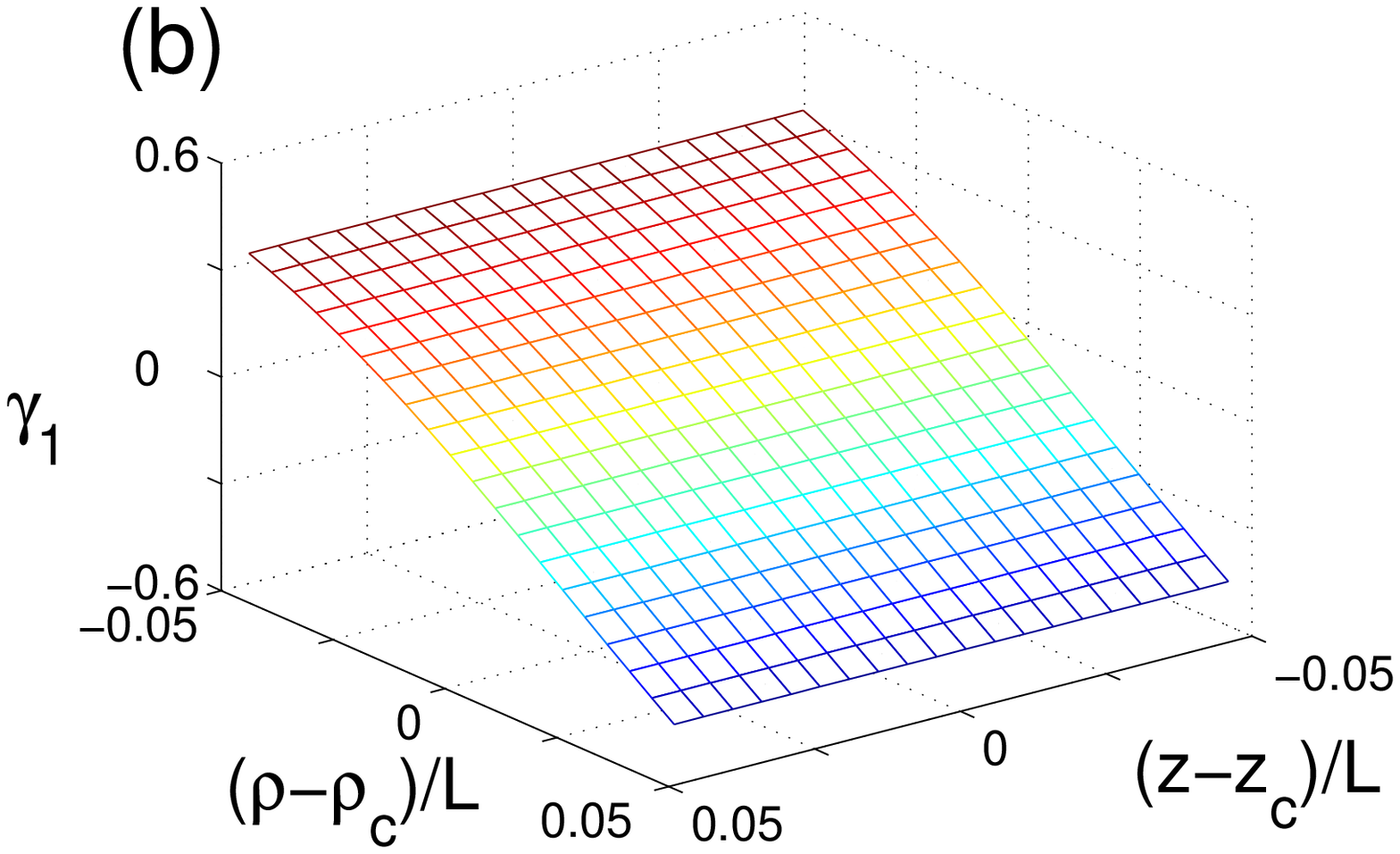}
\caption{(Color online) The spatial distribution of the geometric
phase $\gamma_{1}$ for the storage ring of Refs.
\cite{Rf-classical,Rf-quantum} at (a) $\lambda=1/3$ and (b)
$\lambda=3$. $\eta=\kappa\pi/2$. $B_{\rm rf}^{(0)}=0.08B'L$ and
$B''=10^{-12}B'/L$ are assumed.}
\label{fig6}
\end{figure}

In Fig. \ref{fig6}, we show the spatial distribution of the
geometric phase $\gamma_{1}$ around the trap center with
$\lambda=1/3$ and $\lambda=3$. The fluctuation is found to be
relatively small when $\lambda$ is small or when the rf-field
amplitude $B_{\rm rf}$ is large.

Although not discussed in Refs. \cite{Rf-classical,Rf-quantum},
a ring shaped trap also can be realized if we take
$\eta=-\kappa\pi/2$. The ``effective B-field" $\vec{B}^{\,\rm eff}$
still lies in the $x$-$y$ plane
\begin{eqnarray}
\vec{B}^{\,\rm eff}_{x}(\vec{r})&=&-\frac{B_{\rm
rf}(z)}{2\sqrt{2}}\cos(2\phi)(1-\cos\beta_{s}(\rho,z)),\nonumber\\
\vec{B}^{\,\rm eff}_{y}(\vec{r})&=&\frac{B_{\rm
rf}(z)}{2\sqrt{2}}\sin(2\phi)(1-\cos\beta_{s}(\rho,z)),
\label{BI2pi}
\end{eqnarray}
clearly giving rise to a non-zero solid angle with respect to a
closed path along the storage ring. Therefore, the term
$\gamma^{(\rm eff)}_{F}$ is non-zero in this case. We choose
the eigenstates $|n(\vec{r})\rangle_{s}$ and
$|n(\vec{r})\rangle_{\rm eff}$ as
\begin{eqnarray}
|n(\vec{r})\rangle_{s}&=&\exp[-i\vec{F}\cdot\hat{n}_{s}(\vec{r})\beta_{s}(\rho,z)]|n\rangle_z,\nonumber\\
 |n(\vec{r})\rangle_{\rm
eff}&=&\exp[-i\vec{F}\cdot\hat{n}^{\,\rm
eff}_{\perp}(\vec{r})\beta_{\rm eff}(\rho,z)]|n\rangle_z.
\label{nneff}
\end{eqnarray}
with the unit vector $\hat{n}^{s}_{\perp}(\vec{r})$ in the
$x$-$y$ plane orthogonal to $\vec{B}_{s}(\vec{r})$.
In this case,
the gauge potential $A^{(\phi)}_{n}$ can be expressed as
\begin{eqnarray}
A^{(\phi)}_{n}(\rho,z)=\frac{n}{\rho} \cos\beta_{\rm
eff}(\rho,z)[1+\cos\beta_{s}(\rho,z)].
\end{eqnarray}

\begin{figure}
\includegraphics[height=2.35in]{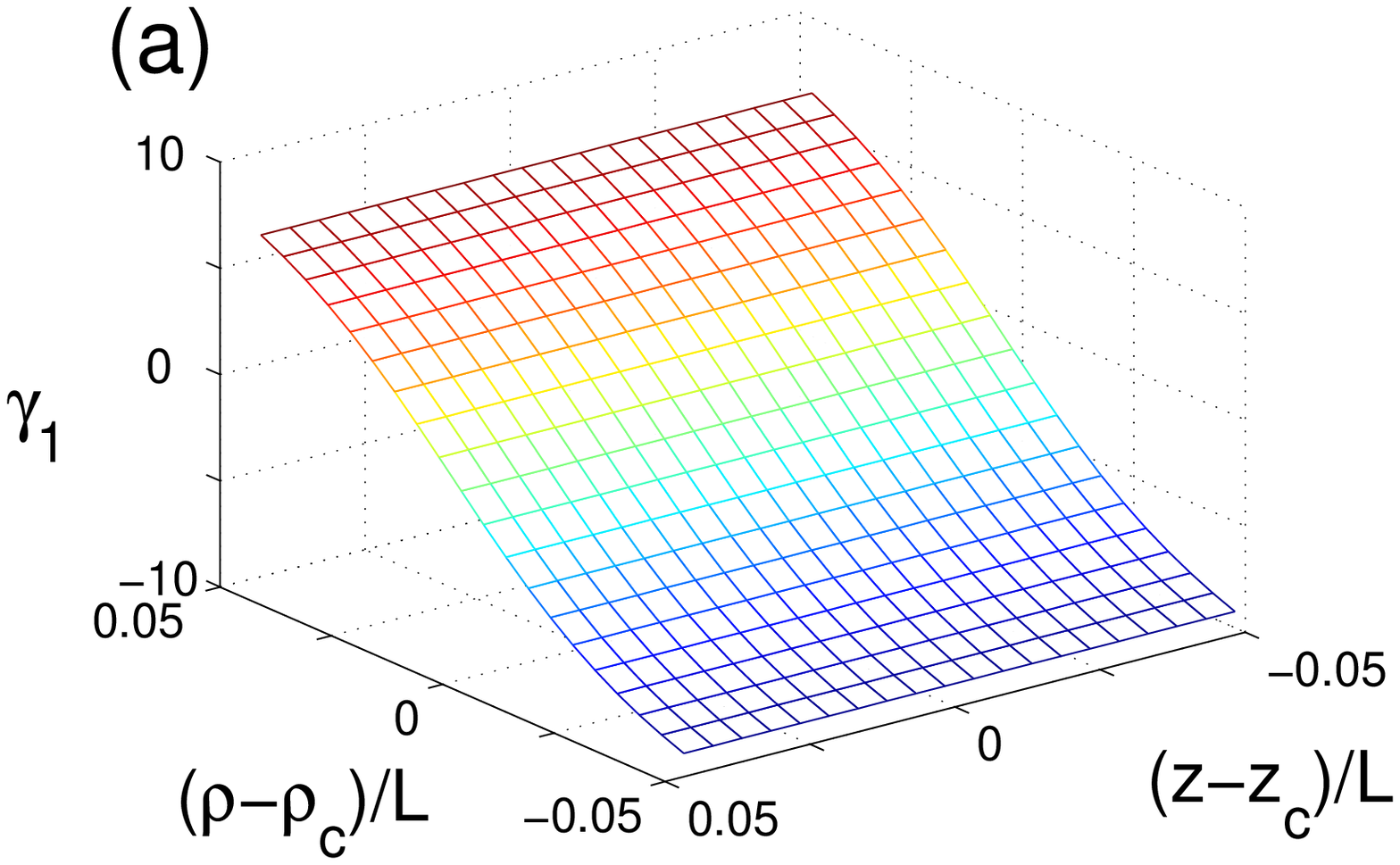}
\includegraphics[height=2.35in]{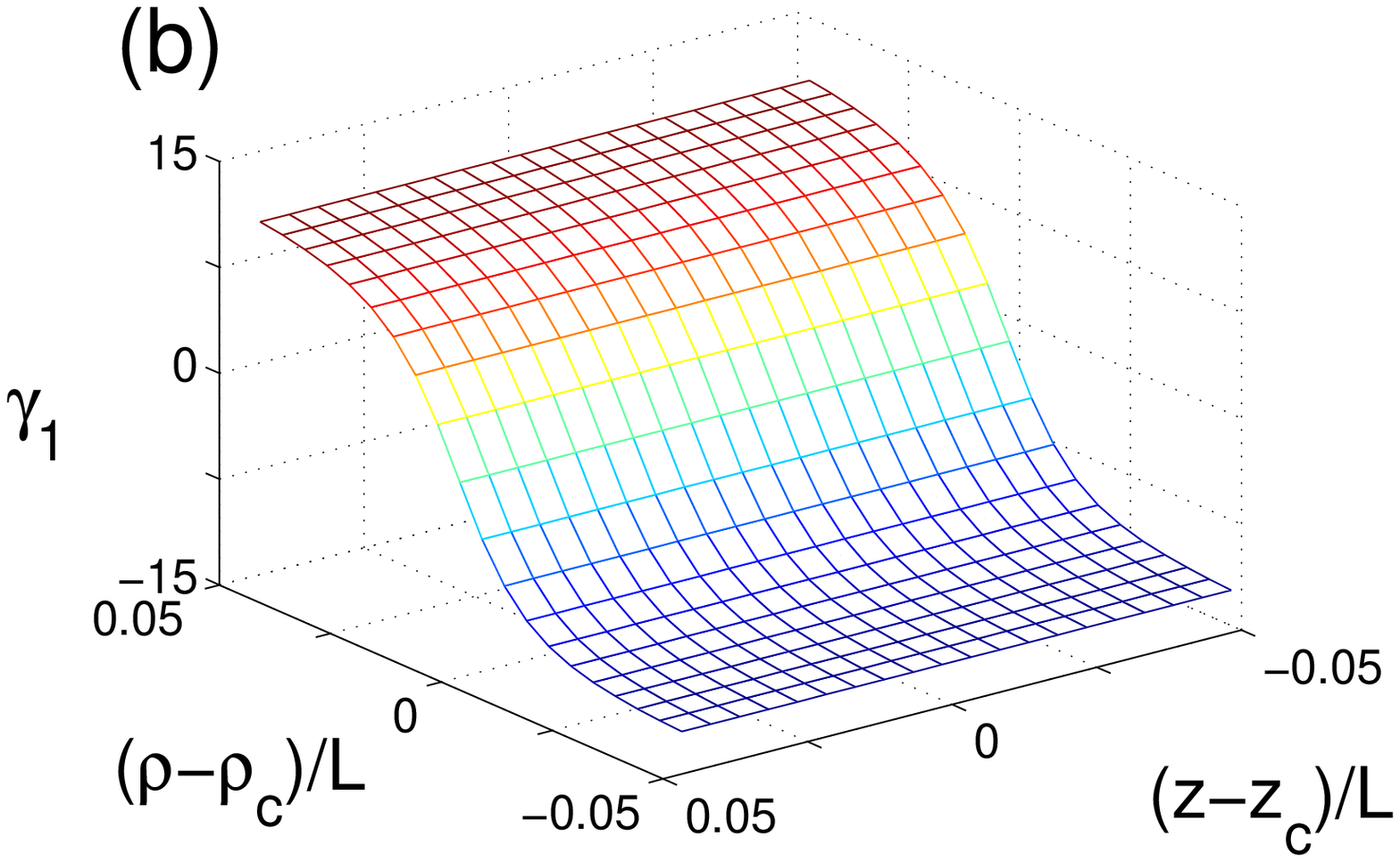}
\caption{(Color online) The spatial distribution of the geometric
phase $\gamma_{1}$ for the storage ring of Refs.
\cite{Rf-classical,Rf-quantum} at (a) $\lambda^{\prime}=3$ and
(b)$\lambda^{\prime}=1/3$. $\eta=-\kappa\pi/2$. $B_{\rm
rf}^{(0)}=0.08B'L$ and $B''=10^{-12}B'/L$ are assumed.}
\label{fig7}
\end{figure}

In Figure \ref{fig7}, we show the fluctuation of the geometric
phase $\gamma_{1}$ for a closed path with a new parameter
\begin{eqnarray}
\lambda'=6\sqrt{2}\frac{\Delta[\vec{r}=0]}{|g_{F}\mu_{B}B_{\rm
rf}^{(0)}|}\, ,
\end{eqnarray}
equal to $3$ and $1/3$. The fluctuation for $\gamma_{1}$ is
found to be much larger than the case of $\eta=\kappa\pi/2$,
which can be explained
by the transverse components $B^{\,\rm eff}_{x,y}$ of the
``effective B-field." Because
$\cos\beta_{s}$ is always close to unity. In the case of $\eta=-\kappa\pi/2$,
$B^{\,\rm eff}_{x,y}$ can take only small positive values.
Therefore, at the minimum of the ARFP trap $\rho=\rho_{0}$
of $|\vec{B}^{\,\rm eff}|$, both $B^{\,\rm
eff}_{z}$ and $B^{\,\rm eff}_{x,y}$ have to be close to zero.
In this case the value for $\cos\beta_{\rm eff}$ becomes a rapidly
changing function of $\rho$ in the region near $\rho_c$.

Our above calculations have obtained analytical expressions of the
geometric phases in an ARFP based storage ring for
$\eta=\pm\kappa\pi/2$. We have further investigated the
fluctuations of the geometric phase for the two cases of
$\eta=\pm\kappa\pi/2$. It seems one benefits from implementing a
Sagnac interferometer in the discussed ARFP storage ring with
$\eta=\kappa\pi/2$ and operating at a relatively large $\lambda$.

Before proceeding onto the concluding section, we will
discuss the geometric phase in an ARFP based
beam splitter created via a double potential
\cite{Rf-classical,Rf-experiment}.
In such an implementation, the
static field $\vec{B}_{s}$ is created from a Ioffe-Pritchard
trap, while the oscillating rf field components are $\vec{B}_{\rm
rf}^{(a)}=B_{\rm rf}[z]\hat{e}_{x}$ and $\vec{B}_{\rm rf}^{(b)}=0$.
By spatially
tuning the amplitude of $B_{\rm rf}$ from zero to a significant
value, in the $x$-$y$ plane, an ARFP can be tuned from a
single well centered near the origin to a double well with two
minimal points at the point with nonzero radius $\rho_{0}$ and
$\phi=0,\pi$. Therefore, a Y-shaped atom beam splitter can be accomplished
when the $B_{\rm rf}[z]$ initially is increased along the $z$-axis
to a large value, and then decreased to zero. In such an
arrangement, the atom beam moving along the $z$ direction can be
separated into two beams that move along the $z$-axis at
$\phi=0,\pi$ for a while, and then can be recombined again into a single
beam.

In the atom interferometer considered above,
both the static field $\vec{B}_{s}$ and the ``effective B-field"
$\vec{B}^{\,\rm eff}$ are limited to the $x$-$z$ plane.
Therefore, for motion along the closed path of the trap bottom,
the solid angle enclosed by the trajectory of
$\vec{B}^{\,\rm eff}$ is zero. Thus, the geometric phase
in (\ref{geometricphase}) can be expressed as
\begin{eqnarray}
\gamma_{n}(t)=-i\int_{0}^{t}\sum_{l}|\,_{\rm eff}\langle
n(\vec{r})|l\rangle_z|^{2}\,_s\langle
l(\vec{r})|\nabla|l(\vec{r})\rangle_{s}\cdot\vec{v} dt'.\ \ \
\end{eqnarray}
We can show that the product $\,_{\rm eff}\langle
n(\vec{r})|l\rangle_z|^{2}\,_s\langle
l(\vec{r})|\nabla|l(\vec{r})\rangle_{s}$ is a function of $\rho_c$
and is independent of $z$. Thus, the geometric phase can be
expressed as an integral of this function with respect to
$\rho_c$, from zero to a large value and then back to zero.
Therefore, the value of the geometric phase would be zero in the
end.

\section{Conclusion}

In this study, we develop theoretical formalisms
for the calculation of the atomic geometric phase
inside an ARFP. We show that, due to the complexity
of the ARFP, the geometric phase depends on the
spatial variation of both the static field and
an ``effective B-field" $\vec{B}^{\,\rm eff}$.
We provide general expressions for the geometric phase
and the corresponding adiabatic gauge potential
in Eqs. (\ref{geometricphase}) and (\ref{An}),
respectively.

To shed light on actual
applications of the atomic geometric phase,
we investigate the distribution of atomic geometric
phases for several proposed or ongoing experiments with ARFP based
storage rings and atom beam splitters. We prove rigorously that the
geometric phase in the center of the storage rings proposed in
Refs. \cite{Rf-ring,Rf-ring2} is always zero. In addition, we
find that in the storage ring of Ref. \cite{Rf-ring}, the spatial
fluctuation of the geometric phase sensitively depends on the
position of the trap center on the ``resonance toroid." In the
proposals of Refs.
\cite{Rf-ring2,Rf-combine,Rf-classical,Rf-quantum}, the
fluctuation for the geometric phase becomes significantly
suppressed when the amplitude $B_{\rm rf}$ of the rf-field is
large. In the proposals of
\cite{Rf-classical,Rf-quantum}, the fluctuations of the geometric phase
 also is suppressed if the
angle $\eta$ is set to be $\kappa 2\pi$. In the beam splitter
realized with the double well potential ARFP
\cite{Rf-classical,Rf-experiment}, the geometric phase is
shown to be zero.

Our work helps to clarify the working principle of
trapping neutral atoms in an ARFP and the validity conditions
for the various approximations involved.
We hope our results will shine new light on the proposed
inertial sensing experiments based on trapped atoms in ARFP.

\begin{acknowledgments}
We thank Dr. T. Uzer and Dr. B. Sun for helpful discussions. This
work is supported by NASA, NSF, CNSF, and the 863 and 973 programs
of the MOST of China.
\end{acknowledgments}

\end{document}